\documentclass[apj,twocolappendix,numberedappendix]{emulateapj}

\usepackage{morefloats}
\usepackage{graphicx}
\usepackage{epsfig}
\usepackage{color}
\usepackage{journals}
\usepackage{amsmath,amssymb,latexsym}

\allowdisplaybreaks

\shorttitle{Santa Barabara Cluster with DISPH}
\shortauthors{Saitoh \& Makino}

\begin{document}

\title{Santa Barbara Cluster Comparison Test with DISPH}

\author{Takayuki \textsc{R.Saitoh}\altaffilmark{1}
        \& Junichiro \textsc{Makino}\altaffilmark{1,2,3}
}
\altaffiltext{1}{Earth-Life Science Institute, Tokyo Institute of
Technology, 2--12--1, Ookayama, Meguro, Tokyo, 152-8551, Japan}
\altaffiltext{2}{Department of Planetology, Graduate School of Science /
Faculty of Science, Kobe University 1-1, Rokkodai-cho, Nada-ku, Kobe, Hyogo,
Japan, 657-8501}
\altaffiltext{3}{RIKEN Advanced Institute for Computational Science,
Minatojima-minamimachi, Chuo-ku, Kobe, Hyogo, 650-0047, Japan}
\email{saitoh@elsi.jp}

\begin{abstract}
The Santa Barbara cluster comparison project \citep{Frenk+1999} revealed that
there is a systematic difference between entropy profiles of clusters of
galaxies obtained by Eulerian mesh and Lagrangian smoothed particle
hydrodynamics (SPH) codes: Mesh codes gave a core with a constant entropy
whereas SPH codes did not.  One possible reason for this difference is 
that mesh codes are not Galilean invariant.
Another possible reason is the problem of the SPH method, which
might give too much ``protection'' to cold clumps because of the unphysical
surface tension induced at contact discontinuities.  In this paper, we apply the
density independent formulation of SPH (DISPH), which can handle contact
discontinuities accurately, to simulations of a cluster of galaxies, and compare
the results with those with the standard SPH. We obtained the entropy core when we
adopt DISPH. The size of the core is, however, significantly smaller than those
obtained with mesh simulations, and is comparable to those obtained with 
quasi-Lagrangian schemes such as ``moving mesh'' and ``mesh free'' schemes.
We conclude that both the standard SPH without artificial conductivity and
Eulerian mesh codes have serious problems even such an idealized simulation,
while DISPH, SPH with artificial conductivity, and quasi-Lagrangian schemes have
sufficient capability to deal with it.
\end{abstract}


\section{Introduction} \label{sec:intro}


\cite{Frenk+1999} conducted a comparison project of numerical simulations of the
formation of a massive cluster of galaxies in a cold dark matter (CDM) universe in
which the results of twelve independent simulation codes were compared. Both
Eulerian mesh codes and Lagrangian smoothed particle hydrodynamics
\citep[hereafter SPH;][]{Lucy1977,GingoldMonaghan1977} codes were used in this
comparison project.  In this project, they took into account only gravity and
hydrodynamics so that the comparison was simplified. This project is known as
the ``Santa Barbara cluster comparison project''. 

Although most of global properties reasonably converged, there was a
systematic difference in entropy profiles of the clusters in simulations with
different methods.  The results of the Eulerian mesh codes showed the central
core in entropy profiles, whereas those of SPH codes did not (see figure 18 in
their paper). Later studies confirmed this tendency
\citep[e.g.,][]{Kravtsov+2002, Ascasibar+2003, Springel2005, Voit+2005,
Mitchell+2009, Vazza2011, Almgren+2013}. \cite{Frenk+1999} interpreted that this
difference might reflect the difference in the treatment of shocks of SPH and
mesh codes.

\cite{Agertz+2007} pointed out that there are ``fundamental differences''
between Eulerian mesh and Lagrangian SPH codes \citep[see
also][]{RitchieThomas2001, Okamoto+2003}.  This difference is originated from
the fact that the standard formulation of SPH (hereafter SSPH
{\footnote{
Following \citet{SaitohMakino2013}, we call ``SSPH'' the SPH
formulation using $m/\rho$, where $m$ and $\rho$ are the mass and density of a
particle, as the discrete volume element for the discretization of the equations.}}
) has difficulties when dealing with contact discontinuities.  Since
SSPH requires the differentiability of the density field, the density around a
contact discontinuity has a large error.  In particular, the density of
particles in the low-density side of the contact discontinuity is overestimated
by a large factor.  This error propagates to the pressure evaluation, resulting
in the {\textit {unphysical surface tension}} and the suppression of fluid
instabilities \citep[See figure 1 in][]{SaitohMakino2013}.

We can solve this problem by using the volume element which does not depend on
the density \citep{RitchieThomas2001, Read+2010, SaitohMakino2013, Hosono+2013,
Hopkins2013, Rosswog2015}, by introducing extra dissipation terms
\citep{Price2008, ReadHayfield2012, Kawata+2013}, by using the integration form
to evaluate the derivatives of physical quantities \citep{garcia-Senz+2012,
Rosswog2015}, by employing the non-standard equation of motion which breaks
Newton's third law \citep{Abel2011}, and by adopting the Godunov SPH
\citep{Inutsuka2002} where physical quantities are smoothed twice
\citep{Cha+2010}. 
\cite{Yamamoto+2015} used a diffusive quantity, $y$, 
instead of the mass and mass density, to formulate the SPH approximation.
Each particle has the $Y$, and from that its density is calculated using the
usual formula to calculate mass density from mass of particles in SPH.

Since all of the SPH codes used in the Santa Barbara cluster comparison project
were based on SSPH, the cuspy entropy profile obtained with SPH codes might be,
at least partly, the artifact caused by the unphysical surface tension. In the
hierarchical structure formation scenario, a cluster of galaxies grows through
mergers of a number of small building blocks which contain cold, low-entropy
gas.  The unphysical surface tension of SSPH tends to protect cold gas clumps in
the building blocks against the fluid instabilities and the ram-pressure
stripping, as demonstrated by the ``blob test'' of \cite{Agertz+2007}.  
As a result, cold clumps might reach the center of the cluster due to the
dynamical friction with keeping their low entropy state, resulting in a cuspy
entropy profile of the cluster.

Some of techniques described above have been applied to the cluster
formation simulations in order to investigate their effect on the central
entropy structure.  \cite{Wadsley+2008} added a diffusion term to the energy
equation which mimics the turbulence mixing
{\footnote {The additional term they used is very close to the
artificial conductivity term proposed by \cite{Price2008}. Formally, the
difference between two methods can be reduced the difference in the adopted
forms of the signal velocity \citep{Price2012}.}}
and performed the Santa Barbara cluster simulations.  They found that the
entropy core could be formed, but the size and depth of the entropy core
depended strongly on the diffusion coefficient and numerical resolutions (see
their figures 12 and 13). 
\cite{Power+2014} carried out a set of comparison tests of the cluster formation
in a $\Lambda$CDM universe with SPH and an adaptive mesh refinement (AMR) codes,
focusing on the formation of the entropy core.  In their comparison tests, they used
not only SSPH but also their new SPH, SPHS \citep{ReadHayfield2012} which
avoids multivalues of physical quantities by artificial diffusion terms with a
higher order dissipation switch and increases the force accuracy by adopting a
higher order kernel with a large number of neighbor particles.  In SPHS runs,
the entropy core was formed in the cluster center and its size was consistent to
that obtained by an AMR code, {\tt Ramses} \citep{Teyssier2002}. 
This good agreement is achieved when the maximum dissipation parameter
$\alpha_{\rm max} \ge 1$ (see their figure 6).  On the other hand, in the case
with $\alpha_{\rm max} = 0.1$, the entropy profile has a core, but it is smaller
and the core entropy is lower than those with converged values. In the case with
$\alpha_{\rm max} = 0$, it is close to that found with SSPH. 
Recently, \cite{Sembolini+2016} reported a new comparison test of a numerical
cluster in a $\Lambda$CDM universe. They showed that SPH codes with the
artificial conductivity (hereafter AC) term and SPHS give entropy profiles
which are close to those given by AMRs (i.e., cored profiles), although they did
not provide the results of the parameter tests.

The agreement with the results of AMR calculations does not guarantee the
validity of the results, since Eulerian schemes have their own limitations.  An
obvious one comes from its Eulerian nature.  In order to integrate the motion
and internal energy of fast-moving cold gas clump, very high accuracy is
required. The internal energy of a gas clump can be many orders of
magnitudes smaller than its kinetic energy in the reference frame of the mesh.
Numerical error comes from the kinetic energy, but it affects the internal
energy of the clump.  Consider the case that a molecular cloud moves in a galaxy
following the galactic potential.  The typical temperature of the cloud is
$\sim~10~{\rm K}$ while the virial temperature which corresponds to the typical
kinematic energy is $\sim 10^6~{\rm K}$ for a galaxy of the Milky Way size. The energy
difference between the thermal and kinetic energies of the cloud becomes about
five orders of magnitude.  Very high accuracy is, thus, required to keep the
structure of the clump, and this requirement in turn results in the requirement
of high spacial resolution and small time step.  
Otherwise, a large error is induced by the loss of significant digits and/or
negative pressure.
A remedy for this problem is the dual energy formulation \citep{Bryan+1995}. In
this formulation, both total ($E_{\rm tot}$) and internal ($E_{\rm th}$)
energies are solved independently. 
Then, either $E_{\rm th}$ or $E_{\rm tot}-E_{\rm kin}$, where $E_{\rm kin}$ is
the kinetic energy, is used to evaluate pressure, depending on the fraction of
the thermal energy.  For the high Mach number fluid, e.g., $(E_{\rm tot} -
E_{\rm kin})/E_{\rm tot} < 10^{-3}$, $E_{\rm th}$ is adopted to evaluate
pressure.  Otherwise, since $E_{\rm tot}\sim E_{\rm kin}$, the loss of
significant digits occurs.

Another difficulty is that the Eulerian formulation is not
the Galilean invariant, which leads to the numerical diffusion.
An example of this problem can be seen in figure 14 of \cite{Tasker+2008}. In
Eulerian codes, an initially hydrostatic cluster decays after 1 Gyr evolution,
in particular in {\tt Enzo} \citep{BryanNorman1997} with {\tt Zeus}
\citep{StoneNorman1992} and {\tt FLASH} \citep{Fryxell+2000}, when a
translational velocity was added.  In principle, by going to higher spatial
resolution, we can reduce the rate of decay by reducing the effect of the
numerical dissipation \citep{Robertson+2010}.  Nevertheless, compared to
Lagrangian codes with the same effective accuracy, the computational cost would
be much higher.
The other way is to adopt higher spatial order schemes, such as
essentially non-oscillatory polynomial interpolation
\citep[ENO;][]{Harten+1987}, weighted ENO \citep[WENO;][]{Liu+1994}, and
constrained interpolation profile \citep[CIP;][]{YabeAoki1991} schemes.
However, these schemes are computationally expensive and it is difficult to
combine high-order schemes with AMR in a consistent way. The latter is serious
because joint interfaces of finer and coarse meshes lose their accurracy
\citep{Li2010} and thus careful treatment is necessary.

Recently, novel numerical schemes have been developed and used in the field of
numerical astrophysics.
\cite{Springel2010} developed a moving mesh scheme which is free from the above
problem of the Eulerian mesh codes.  In \cite{Springel2010}, he carried out
simulations of the Santa Barbara cluster with ``{\tt AREPO}'', an implementation
of his moving mesh scheme, and reported that the entropy core was formed.
The size of the core is smaller than those obtained by AMR simulations and the
entropy at the core is lower than those obtained by AMR simulations.
\cite{Hopkins2015} developed a new code ``{\tt GIZMO}'' which implemented a
meshless method and found that in cluster simulations with {\tt GIZMO} entropy
cores similar to those of {\tt AREPO} is formed. 
These results are quite encouraging. 
However these two codes use similar approaches, such as quasi-Lagrangian
nature using Riemann solvers. Further tests with different schemes are
important.

In this paper, we report the results of Santa Barbara cluster simulations
obtained with the density independent formulation of SPH (DISPH)
\citep{SaitohMakino2013}.  Since DISPH is free from the unphysical surface
tension, we can expect that it gives better results compared to SSPH.  In
addition, since DISPH do not introduce artificial dissipations, we can expect
that the result is not affected by free parameters.

The structure of this paper is as follows. In \S \ref{sec:DISPH} we give a brief
review of the formulation of SSPH and DISPH and their advantages and
disadvantages.  We then describe the initial condition and our numerical methods
in \S \ref{sec:method}.  The comparison is carried out in \S \ref{sec:results}
and discussion is given in \S \ref{sec:discussion}.

\section{Differences between SSPH and DISPH} \label{sec:DISPH}

In the SPH method, all physical quantities are evaluated using the kernel
approximation over the field discretized by particles.
The kernel approximation for a physical quantity $f$ is given by 
\begin{equation}
f(\boldsymbol x) = 
\int_{-\infty}^{\infty} f(\boldsymbol x') W(\boldsymbol x'-\boldsymbol x,h) d \boldsymbol x',
\label{eq:int_f}
\end{equation}
where $\boldsymbol x$ and $\boldsymbol x'$ are position vectors, $W$ is a
function with a compact support called the kernel function, and $h$ is the
kernel size.  We approximate this spatial integration by the summation over
particles expressed as
\begin{equation}
f_i = \sum_j f_j W(\boldsymbol x_j-\boldsymbol x_i,h_i) V_j, \label{eq:sum_f}
\end{equation}
where $f_i$ and $f_j$ are the values of $f$ at the positions of particles $i$
and $j$, respectively, and $V_j$ is the volume element associated with
particle $j$.  In SSPH, we use
\begin{equation}
V_j = \frac{m_j}{\rho_j}, \label{eq:SSPH:voluem_element}
\end{equation}
where $m_j$ and
$\rho_j$ are the mass and density of particle $j$, respectively.  From Eqs.
\eqref{eq:sum_f} and \eqref{eq:SSPH:voluem_element}, by substituting $\rho$ into
$f$, we have 
\begin{equation}
\rho_i = \sum_j m_j W(\boldsymbol x_j-\boldsymbol x_i,h_i). \label{eq:SSPH}
\end{equation}

Note that Eq. \eqref{eq:SSPH} gives the {\it smoothed} estimate of the density
field. Thus, the density near a contact discontinuity suffers a very large error.
Consider the case that we evaluate the density around a contact discontinuity
using Eq \eqref{eq:SSPH}.
The density of a particle at the low-density side is over-estimated, while that
of a particle at the high-density side is under-estimated.  The absolute amount
of the error is similar at the both sides. The relative error is much larger at
the low-density side, simply because the true density is smaller. This large
overestimate of the density results in equally large overestimate of the
pressure for low-density particles near the contact discontinuity, which then
causes a strong repulsive force.  This repulsive force works as an unphysical
surface tension \citep[e.g.,][]{RitchieThomas2001, Okamoto+2003, Agertz+2007,
Read+2010}. 
If the distribution of the internal energy is sufficiently smooth,
this unphysical surface tension does not show up even with the smoothed density.
The AC term introduced by \cite{Price2008} automatically reduces the pressure
jump by spreading the internal energy.
{\footnote {
Although AC can eliminate the unphysical surface tension, we consider that it is
not always an adequate solution. For instance, AC has difficulty dealing with a
fluid system with different chemical compositions. See appendix \ref{sec:AC}.}}
A switch is always used to avoid unnecessary diffusion.

In DISPH, unlike SSPH, the pressure (or its arbitrary function) is used to evaluate the
volume element \citep{SaitohMakino2013, Hopkins2013, Hosono+2013, Rosswog2015}.
The pressure field is smooth everywhere, except at the shock front. In SPH, the
shock is handled by the artificial viscosity (hereafter AV) which smooths the velocity field
and prevents multivalues.  As a by-product, we have a smoothed pressure
distribution in SPH simulations, even at the shock front.  Therefore, we can
expect that the problem of SSPH at the contact discontinuity is solved if we use
a volume element based on the pressure. This is the basic idea of DISPH.  The
simplest form of the fundamental equation used in DISPH can be obtained by using
the volume element $V_j = U_j/q_j$, where $q_j$ and $U_j$ are the energy density
and the internal energy of particle $j$, respectively.  Substituting $q$ into
$f$ in Eq. \eqref{eq:sum_f} and using the energy density based volume element,
we have
\begin{equation}
q_i = \sum_j U_j W(\boldsymbol x_j-\boldsymbol x_i,h_i). \label{eq:DISPH}
\end{equation}
Since the pressure is smooth at
contact discontinuities and even at the shock front, we can obtain smoothed
values without large errors.  Note that, if one uses DISPH for a non-ideal gas,
one needs to use pressure directly instead of the energy density
\citep{Hosono+2013}, because it is no longer smooth.
We note that this formulation itself does not introduce any dissipation.

In DISPH, as a trade-off for using of the smoothed pressure for the
formulation, the density distribution has a jump at a contact discontinuity if
the internal energy distribution there is not smooth
\citep[see figure 2 in][]{SaitohMakino2013}.  
While this jump does not affect the motion of particles, it is necessary to be
aware of this ``unphysical'' density jump when one uses this density to evaluate physical
quantities, such as radiative cooling.  As is noted by
\cite{SaitohMakino2013} and \cite{Hopkins2013}, the simplest solution is to use
the smoothed density.  

In the same manner as SSPH, DISPH can conserve the total energy, linear and
angular momentum.  If we use a variation principle for the derivation of SPH
equations \citep{SpringelHernquist2002}, we can obtain a set of conservative SPH
equations \citep{Hopkins2013}. The equations in this derivation have the
so-called ``grad-$h$'' term.  Moreover, if we carefully constructs energy and
momentum equations, we can obtain another set of the conservative equations
which do not involve the grad-$h$ term \citep[see section 5.6
in][]{SaitohMakino2013}.  Here we adopt DISPH with the grad-$h$ term, which is
more robust under the strong shock. In appendix
\ref{sec:method:TimeIntegration}, we show the result of the three-dimensional
collapse test \citep{Evrard1988}. The relative energy error from the beginning
and the end of the simulation is $\sim 0.2$\%, which is comparable to the
previous studies \citep[e.g.,][]{Springel+2001,Springel2005}.

\section{Method} \label{sec:method}

\subsection{Initial condition}

The initial condition we used here is basically the same as that used in
\cite{Heitmann+2005}. The volume of their initial condition is $64^3~{\rm
Mpc^3}$ in the comoving space and the initial redshift is $z = 63$.  The
$3\sigma$ overdensity is imposed at the center of the volume so that a massive
cluster will be formed at the center.  There are two set of initial conditions
with different resolutions: $128^3$ and $256^3$ particles.  The cosmology is the
standard CDM and the Hubble parameter is $H_0 = 100~h~{\rm km~s^{-1}~Mpc^{-1}}$
where $h = 0.5$.

We integrated the formation and evolution of the Santa Barbara cluster with a
vacuum boundary condition in the physical space.  We cut the central spherical
volume with the radius of $32~{\rm Mpc}$ in the comoving space and remapped it
to the physical space at $z=63$.  Then, we imposed the Hubble flow corresponding
to the initial redshift.  Following \cite{Frenk+1999}, we assumed $\Omega_{\rm
DM} = 0.9$ and $\Omega_{\rm baryon} = 0.1$.  Since the original initial
conditions of \cite{Heitmann+2005} are for DM only simulations, we divided each
particle into two particles which have the same position and velocity of the
original particle but have different masses, i.e., that one has $90\%$ and the
other has $10\%$ of the original particle mass, for simulations involving both
DM and gas.

The number of particles, the mass of particles and the spatial resolutions
(softening lengths for DM and baryon: $\epsilon_{\rm DM}$ and $\epsilon_{\rm
baryon}$) are summarized in table \ref{tab:parameters}.  For both of low- and
high-resolution models, we performed two runs, one with the standard SPH, and
the other with DISPH.  We carried out five high-resolution SSPH runs involving
AC.  These five runs adopted different values for the maximum dissipation
parameter.  In addition, two extra models which combine the low mass and the
high spatial resolutions (Mixed 1) and the high mass and the low spatial
resolutions (Mixed 2) are adopted in order to investigate the effect of the
resolutions further. We also carried out two runs with SSPH and DISPH for each
model. Thus, in total we performed thirteen runs.

\begin{table*}[htb]
\centering
\caption{Number and mass of particles and spatial
resolutions.}\label{tab:parameters}
\begin{tabular}{rccccc}
\hline
\hline
Run type&Number&DM mass & Baryon mass& $\epsilon_{\rm DM}$ & $\epsilon_{\rm Baryon} $\\
\hline
Low resolution& $2202482~(1101241\times2)$ &$7.8\times10^9~M_{\odot}$&
$8.8\times10^{8}~M_{\odot}$&$20~{\rm kpc}$&$10~{\rm kpc}$\\
High resolution&$17619230~(8809615\times2)$&$9.6\times10^{8}~M_{\odot}$&
$1.1\times10^{8}~M_{\odot}$&$10~{\rm kpc}$& $5~{\rm kpc}$\\
\hline
Mixed resolution 1 &$17619230~(8809615\times2)$&$9.6\times10^{8}~M_{\odot}$&
$1.1\times10^{8}~M_{\odot}$&$20~{\rm kpc}$& $10~{\rm kpc}$\\
Mixed resolution 2 & $2202482~(1101241\times2)$ &$7.8\times10^9~M_{\odot}$&
$8.8\times10^{8}~M_{\odot}$&$10~{\rm kpc}$&$5~{\rm kpc}$\\
\hline
\end{tabular}\\
\end{table*}

\subsection{Numerical techniques}\label{sec:method:numerical}

We used a parallel $N$-body/SPH code {\tt ASURA} \citep{Saitoh+2008,
Saitoh+2009} for this experiment.  The gravitational interactions between
particles were calculated by the tree method \citep{BarnesHut1986}. The tree
with GRAPE method \citep{Makino1991TreeWithGRAPE} was used.  The parallelization
method of the gravity calculation was that proposed by \citet{Makino2004}.  We
adopted the symmetrized Plummer potential and its multipole expansion
\citep{SaitohMakino2012} so that we can calculate the gravitational interactions
between particles with different values of softening lengths by a single tree.
The opening angles for the ordinary three dimensions and softening lengths were
set to $0.5$ following \cite{SaitohMakino2012}.  The phantom-GRAPE library was
used for the calculation of particle-particle interactions in order to
accelerate the calculation \citep{Tanikawa+2013}.

The equation of state of the ideal gas with the specific heat ratio $\gamma =
5/3$ was used.  In order to handle shocks, the AV term
proposed by \cite{Monaghan1997} was used.  In addition to this, we used an
AV switch proposed by \cite{MorrisMonaghan1997} in which the
value of the AV parameter, $\alpha$, changes in the range of $0.1-1$.
Following \cite{Rosswog2009}, a modification which accelerates the increase of
$\alpha$ at lower $\alpha$ regime and slows it at higher $\alpha$ regime was
also used.  The Balsara limiter \citep{Balsara1995} was adopted in order to
reduce the shear viscosity.
We employed the Wendland C4 kernel which is free from the paring instability
\citep{Wendland1995, DehnenAly2012}. The number of neighbor particles was kept
within $128\pm8$.  This higher order kernel with a larger number of neighbor
particles can reduce the so-called E0 error \citep{Read+2010}.

The time integration was carried out by a second order scheme
(see appendix \S \ref{sec:method:TimeIntegration} for details).
The individual and
hierarchical time-step methods were used \citep{McMillan1986,
Makino1991IndividualTimeStep}.  For SPH particles, the time-step limiter, which
enforces the time-step difference among neighboring particles to be small enough
to follow strong shocks, was used \citep{SaitohMakino2009}. The FAST scheme,
which allows each particle to have different time-steps for gravitational and
hydrodynamical interactions and integrates these two interactions independently,
was also adopted \citep{SaitohMakino2010}.
When we used the FAST scheme, the computational time (wall-clock
time) became half of what it was without FAST (we used 128 CPU cores of Cray XC30 in
this comparison). This gain is almost the same as that reported in
\cite{SaitohMakino2010}, although simulation setups in both runs were completely
different.  

\subsection{Detection of the Santa Barbara Cluster} \label{sec:results:Detection}

Following previous studies \citep{Frenk+1999, Heitmann+2005}, we define a dense
region which is formed at the center of the simulation volume and with the mean
density is 200 times higher than the background density as the ``Santa Barbara
cluster''. We express the radius where the mean density becomes 200 times of the
background density as $R_{200}$ and call it the virial radius.  

The procedure to find the center and size of the Santa Barbara cluster is as
follows.  First, we find a particle of which total energy is the lowest and
adopt it as the center of the cluster of galaxies. Then, we sort all of
particles based on the distance from the center of the cluster in ascending
order.  We calculate the mean density from the center to the outer part
following the particle order, and we halt this operation when the mean density
becomes less than 200 times of the background density. We regard the distance
from the center to the particle position where we halt this operation as the
virial radius.  This simple strategy works well for the Santa Barbara test in
which a single massive cluster is formed near the center of the simulation
volume.

\subsection{Definitions of the Density Center and the Core Radius} \label{sec:method:Center}

In order to draw radial profiles, we use a density weighted center instead of
the halo center since the peak of the baryon distribution does not always
coincide with the position of particle with the lowest total energy.  We use the
following definition as the density weighted center
\citep{vonHoerner1960, vonHoerner1963}:
\begin{equation}
\boldsymbol x_{\rm c} = \frac{\sum_{j} \rho_j \boldsymbol x_j}{\sum_j
\rho_j}, \label{eq:DensityCenter}
\end{equation}
where we only take care of the gas particles in the virial radius.  
For this density $\rho_j$, we just used the SSPH-estimated density.  The kernel
function and the number of neighbor particles are the same as those described in
\S \ref{sec:method:numerical}.

\section{Results} \label{sec:results}

In this section, we compare the properties of the Santa Barbara cluster obtained
by simulations with SSPH and DISPH, and then investigate the origin of the
difference in the entropy profile.  We first describe the global properties of
the Santa Barbara cluster in \S \ref{sec:results:GlobalProperties}.  The
density, temperature and entropy distributions in the clusters obtained by
simulations with SSPH and DISPH at $z=0$ are compared in \S
\ref{sec:results:Snapshots}.  In \S \ref{sec:results:RadialProfile}, we compare
radial profiles.  We investigate how the central entropy core/cusp formed in \S
\ref{sec:results:Entropy:Formation}.  
In \S \ref{sec:results:AC}, we investigate the contribution of the AC term
for the cluster entropy profile.

In order to avoid confusion, we use the label ``SPH'' to indicate the results of
SSPH on figures and ``DISPH'' for the results of DISPH.

\subsection{Global Properties of the Santa Barbara Cluster at $z=0$}
\label{sec:results:GlobalProperties}

The average of virial radii of the Santa Barbara clusters of the four runs (the
high- and low-resolution runs with SSPH and DISPH) is $R_{200}\sim2.9~{\rm
Mpc}$.  This radius is somewhat larger than those obtained in previous studies
\citep{Frenk+1999, Heitmann+2005}. The total mass within $R_{200}$ is $\sim 1.4
\times 10^{15}~M_{\odot}$ which is $16\%$ larger than those obtained by
\cite{Heitmann+2005}.  These differences are probably due to the difference in
the treatment of the boundary condition.  We used the open boundary condition,
while all runs in \cite{Heitmann+2005} adopted the periodic boundary condition.
Since the radius of the initially imposed overdensity region is $10~{\rm Mpc}$,
the cut-off radius of $32~{\rm Mpc}$ would be insufficient to remove boundary
effects.  Nonetheless, as we see later, the central region of the cluster does
not seem to be affected by the boundary condition.  In the following, we call
the Santa Barbara cluster just ``cluster'', for brevity.

We found that the clusters obtained by DISPH contain more gas than those
obtained by SSPH. The gas fractions of the cluster with DISPH are $0.097-0.098$
which are close to the results of high resolution mesh codes in
\cite{Frenk+1999} and also the cosmic averaged value of $0.1$ for this model. On
the other hand, those of SSPH are $0.093-0.094$ which are close to those
of SSPH runs in \cite{Frenk+1999}, $\sim 0.09$.

\subsection{Snapshots of the Santa Barbara Cluster at $z=0$} 
\label{sec:results:Snapshots}

\begin{figure*}[htb]
\centering
\epsscale{1.0}
\plotone{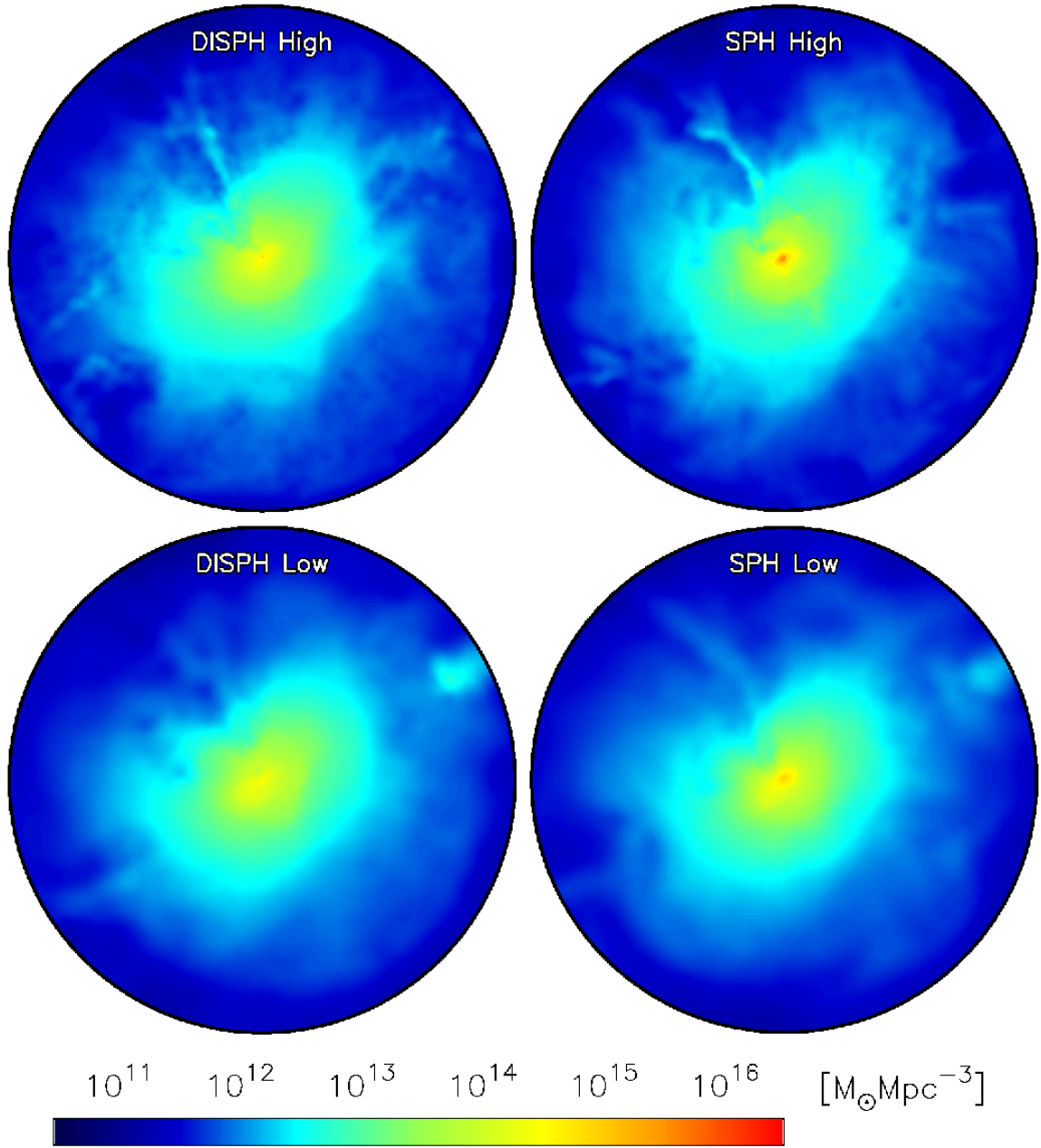}
\caption{Density maps of four runs at $z = 0$.  The center of the cluster is
adopted as the center of the coordinates.  The radius of the plot region in
each density map is corresponding to the virial radius.  In the virial radius,
we paved meshes whose sizes are $\sim 60~{\rm kpc}^2$ and then evaluated the
density of each mesh center using Eq. \ref{eq:SSPH} where the neighbor number is
$128\pm8$.  We did not apply any smoothing filter for these maps.
\label{fig:Virial:Density}
}
\end{figure*}

Figure \ref{fig:Virial:Density} shows the gas density maps of the four runs
within $R_{200}$ at $z=0$.  These density maps are not very different.  However,
when we inspect these maps carefully, we find several differences.  First, we
can see that the central densities in runs with SSPH are higher than those with
DISPH.  The density of the innermost regions of the clusters with SSPH is one
order of magnitude higher than that with DISPH.  Second, for both of SSPH and
DISPH runs, the central density of the cluster is higher for runs with high mass
resolution (large number of particles).

\begin{figure*}[htb]
\centering
\epsscale{1.0}
\plotone{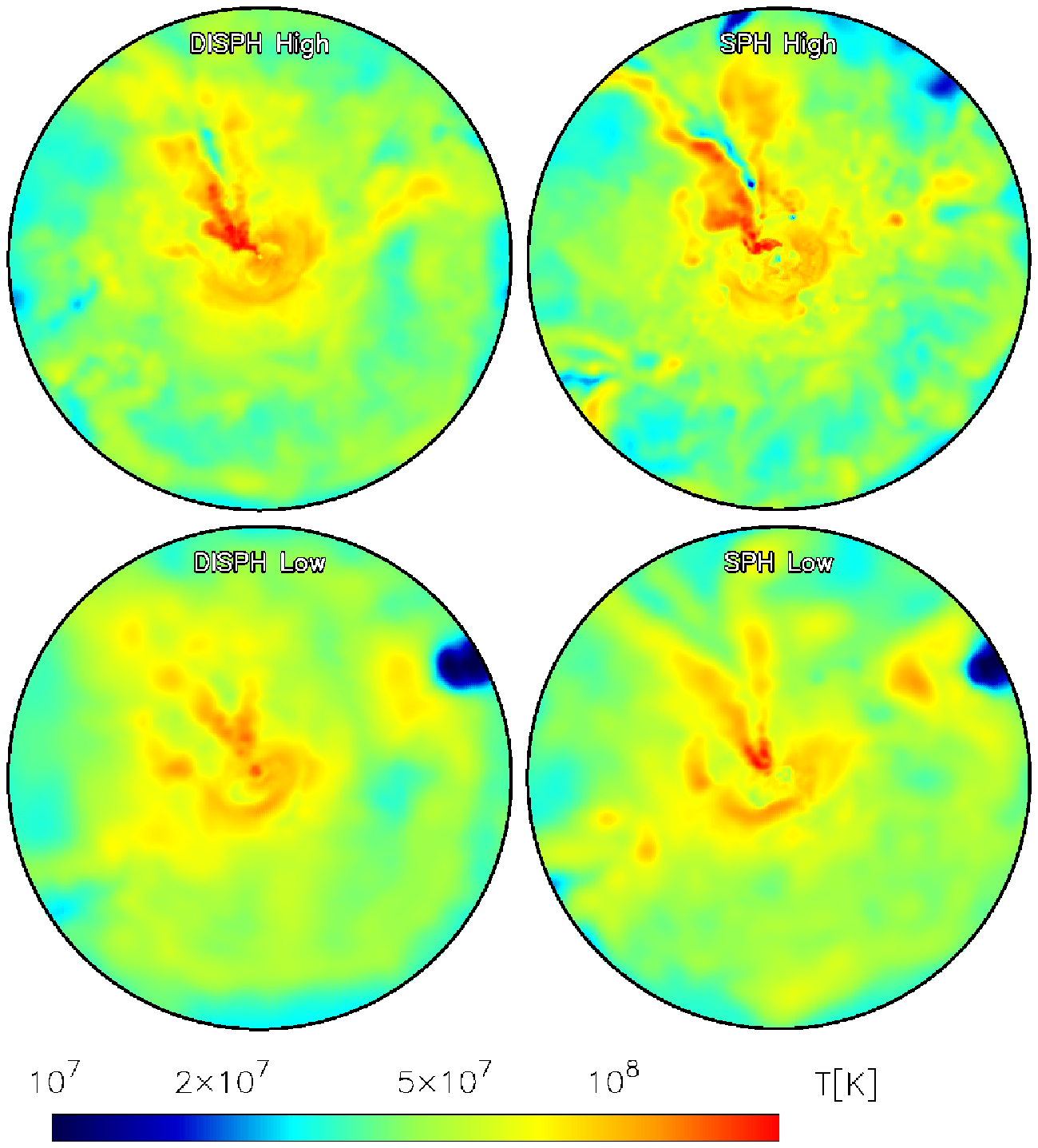}
\caption{The same as figure \ref{fig:Virial:Density} but for temperature.
\label{fig:Virial:Temperature}
}
\end{figure*}

Figure \ref{fig:Virial:Temperature} shows the temperature distribution within
the cluster of four runs. We can see that the clusters of SSPH runs contain more
small-scale substructures compared to those of DISPH runs.

\begin{figure*}[htb]
\centering
\epsscale{1.0}
\plotone{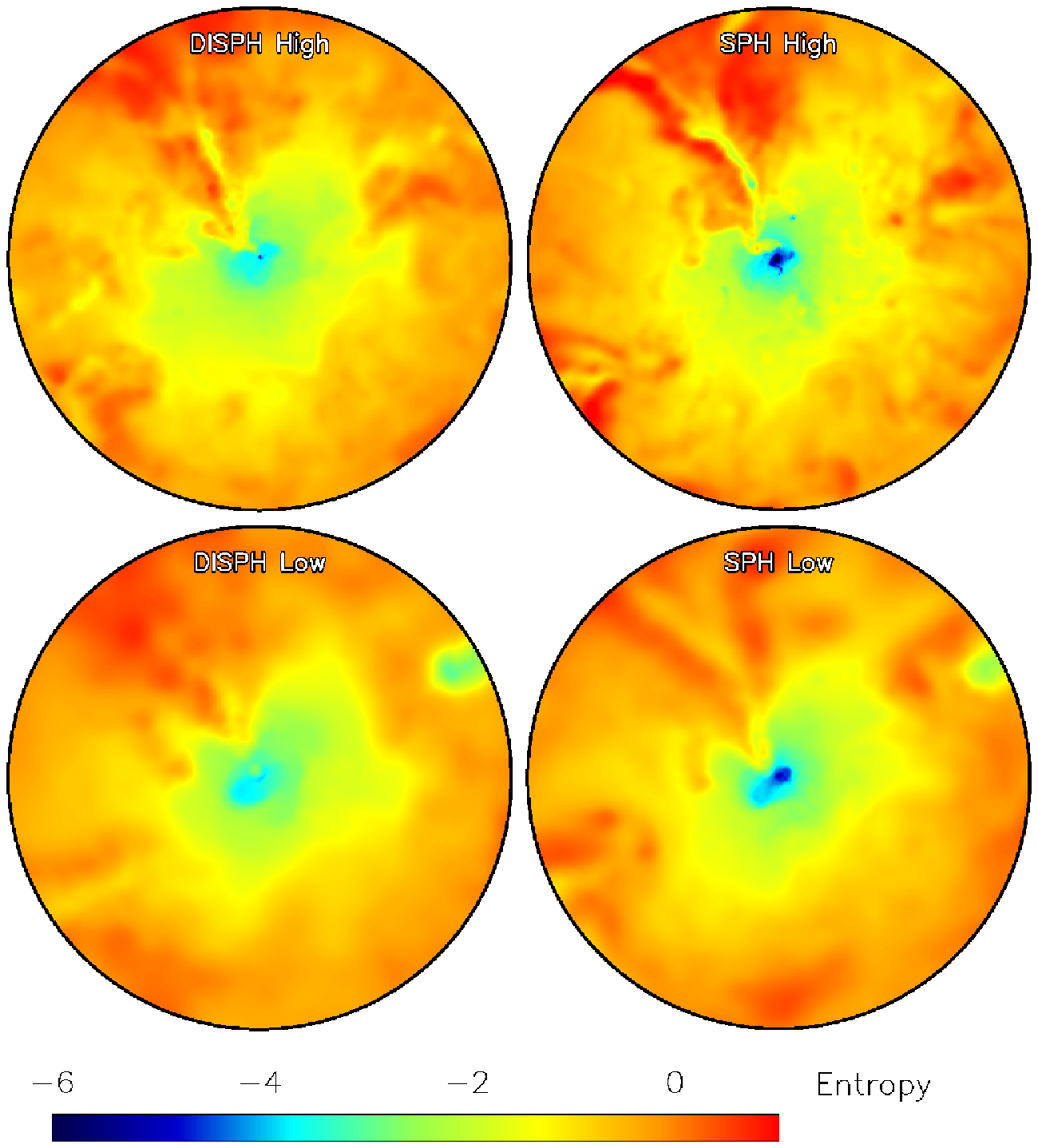}
\caption{The same as figure \ref{fig:Virial:Density} but for entropy.
\label{fig:Virial:Entropy}
}
\end{figure*}

Figure \ref{fig:Virial:Entropy} shows the entropy distributions; the definition
of the entropy we adopted in this paper is \begin{equation} s\equiv\ln
(T/\rho^{2/3}), \end{equation} where $T$ and $\rho$ are the temperature in the
unit of Kelvin and the density in the unit of ${\rm {M_{\odot}~Mpc^{-3}}}$,
respectively, and this definition is the same as that used in \cite{Frenk+1999}.
It is obvious that the values of the central entropy in SSPH runs are much lower
than those in DISPH runs. The clusters in DISPH runs have no or a little amount
of gas with entropy below $-4$, while the clusters in SSPH runs have significant
amounts of such gas at the center. It is also clear that the entropy
distributions of the cluster in DISPH runs are smoother than those of clusters
in SSPH runs.  In other words, the entropy distribution indicates that the gas
in the clusters of SSPH runs is not well mixed.

\subsection{Radial Profiles of the Santa Barbara Cluster at $z=0$}
\label{sec:results:RadialProfile}

\begin{figure*}[htb]
\centering
\epsscale{1.0}
\plotone{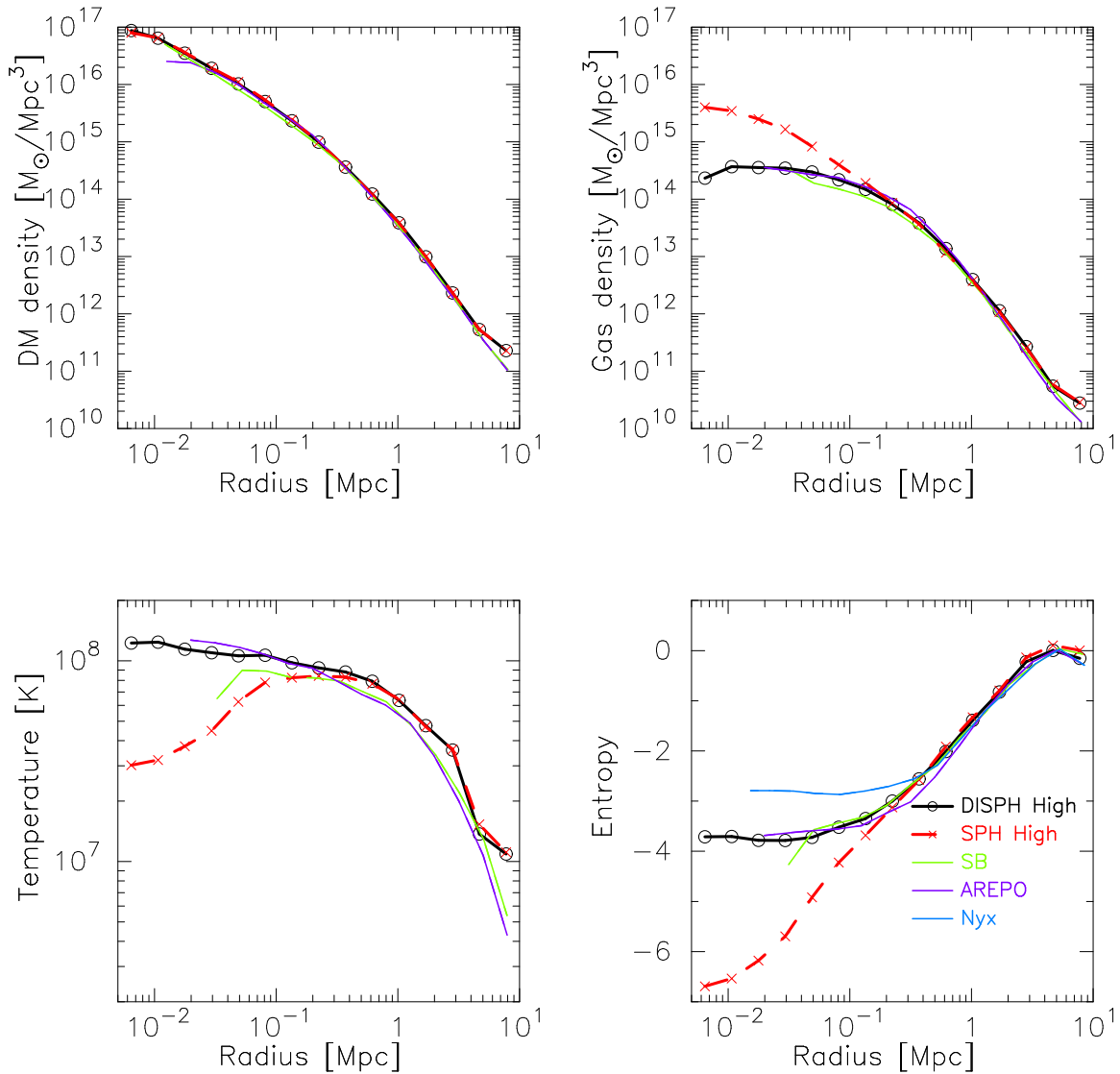}
\caption{Radial profiles of the cluster at $z=0$ for the high
resolution runs and several reference results.  Top-left, top-right, bottom
left, and bottom right panels show the DM density, gas density, temperature,
and entropy profiles, respectively.  Solid black and dashed red curves are the
profiles with the high resolution DISPH and SSPH.  Solid, thin curves with
light-green, purple, regatta-blue are the averaged profiles of
\cite{Frenk+1999} (from figure 18 in their paper), the profiles of the moving
mesh code {\tt AREPO} \citep{Springel2010} (from figure 45 in his paper. We
adopt the result labeled $128^3$), and the profiles of an AMR code {\tt Nyx}
\citep{Almgren+2013} (from figure 7 in their paper), respectively.  The AMR
result is only shown in the entropy panel.  \label{fig:radialprofiles}
}
\end{figure*}

Figure \ref{fig:radialprofiles} shows the radial profiles of the 
cluster at $z=0$. In this figure, we used two high-resolution results.
In this plots, ``SB'' denotes the averaged profiles in \cite{Frenk+1999},
``AREPO'' is the result of a moving mesh code {\tt AREPO} by
\cite{Springel2010}
{\footnote{
We note that the plot data of ``AREPO'' is based on the run with
the ``entropy-energy formalism'', where the value of the entropy is completely
conserved in all cells whose Mach numbers are less than $1.1$.  As expected,
with this formalism the entropy generation is slightly suppressed
\citep{Sijacki+2012}.  On the other hand, \cite{Hopkins2015} pointed out that
spurious heating from Riemann solver errors affects an entropy profile when this
formalism turned off.  Hence, we adopt the original data shown in
\cite{Springel2010}.
}}
, and ``Nyx'' denotes the result of an AMR code {\tt Nyx} by \cite{Almgren+2013}.
{\footnote { {\tt Nyx} is a $N$-body/gas dynamics code designed for
large scale cosmological simulations. In this code, the fluid evolution is
solved by a finite volume method with a block structured AMR. The dual energy
formulation is used. The Riemann solver is used for the evaluation of fluxes.
According to these features, {\tt Nyx} is one of good representatives of the
state-of-the-art AMR codes. The entropy profile of the Santa Barbara cluster
with {\tt Nyx} is almost identical to those with {\tt Enzo} \citep{Bryan+1995}
and {\tt ART} \citep{Kravtsov+2002}. }}
We note that these data were taken from figures in their papers: figure 18 in
\cite{Frenk+1999} for SB, figure 45 in \cite{Springel2010} for AREPO (we adopt
the result labeled $128^3$), and figure 7 in \cite{Almgren+2013} for Nyx.  We
also note that the entropy profile with {\tt GIZMO} reported by \cite{Hopkins2015} is
comparable to that with {\tt AREPO}.

The density profiles of DM (the top-left panel of figure \ref{fig:radialprofiles})
of all four runs agree very well.  The central density of gas of the SSPH run is
significantly higher than those of the other three runs. Here, ``SB'', ``AREPO''
and our DISPH results seem to agree well. Note, however, that the ``SB'' plot is
the average of 7 SPH and 5 mesh-based scheme runs, and SPH models in
\cite{Frenk+1999} gave the result similar to our SSPH run, while the mesh runs
gave a central density even lower than AREPO/DISPH results.  Thus, the
agreement between our DISPH result and the AREPO result is rather good, whereas
all SSPH runs gave high central densities and AMR runs, lower density.

From the temperature profile (the left-bottom panel of figure
\ref{fig:radialprofiles}) and density profile (the top-right panel) of the gas
component, we can see that the standard SPH gives a central cusp with the
temperature decreasing inward, while AREPO and DISPH give constant-density,
constant-temperature cores.  Mesh runs in \cite{Frenk+1999} also gave
constant-density cores, but their cores are systematically larger than those
obtained by AREPO and our DISPH.  The averaged ``SB'' profile of the temperature
shows a sign of central decrease, since it is the average of SPH (decreasing)
and mesh (increasing) profiles. We can see the same tendency in the entropy
profile (the bottom-right panel).

\begin{figure}[htb]
\centering
\epsscale{1.0}
\plotone{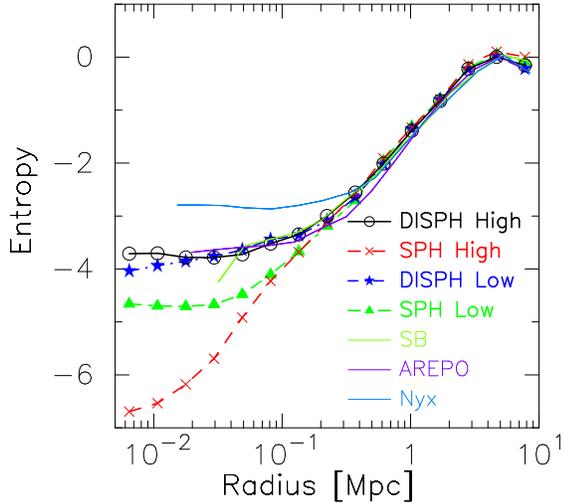}
\caption{Comparison of radial entropy profiles of the cluster at $z=0$. The
solid (black) and dashed (red) curves are the high resolution runs with DISPH
and SSPH, whereas the dot-dashed (blue) and dot (green) curves are the low
resolution runs with DISPH and SSPH. The other thin, solid curves are the same
as those in the top-right panel in figure \ref{fig:radialprofiles}.
\label{fig:Entropy:Comparison}
}
\end{figure}

In figure \ref{fig:Entropy:Comparison},  we plot the entropy profiles of our
low-resolution runs, as well as those of our high-resolution runs and three results
from the literature (``SB'', ``AREPO'' and ``Nyx''). We can see that for DISPH
runs, the difference in resolution does not make much difference in the
entropy profile; the entropy profiles with DISPH are consistent with that
obtained by a moving mesh code {\tt AREPO}. 
On the other hand, our low-resolution SSPH run gave a central
entropy significantly higher than that of our high-resolution SSPH run.  Our
low-resolution SSPH result is similar to those of the SSPH simulations in
\cite{Frenk+1999} and \cite{Springel2010}. Thus, it is clear that the ``core''
in these previous SSPH results is due to the limitations in the numerical
resolution. 

From this figure, we find that the central entropy in the high-resolution DISPH
run is slightly higher than that in the low-resolution DISPH run, although we
saw the opposite sense from the entropy maps (figure \ref{fig:Virial:Entropy}).
This is because there is an offset of $\sim 50~{\rm kpc}$ between the position
of the entropy minimum and the plot center (the density weighted center) in the
high-resolution DISPH run, and it makes the rather flatter central entropy
profile. Thus, there is no discrepancy.

\begin{figure}[htb]
\centering
\epsscale{1.0}
\plotone{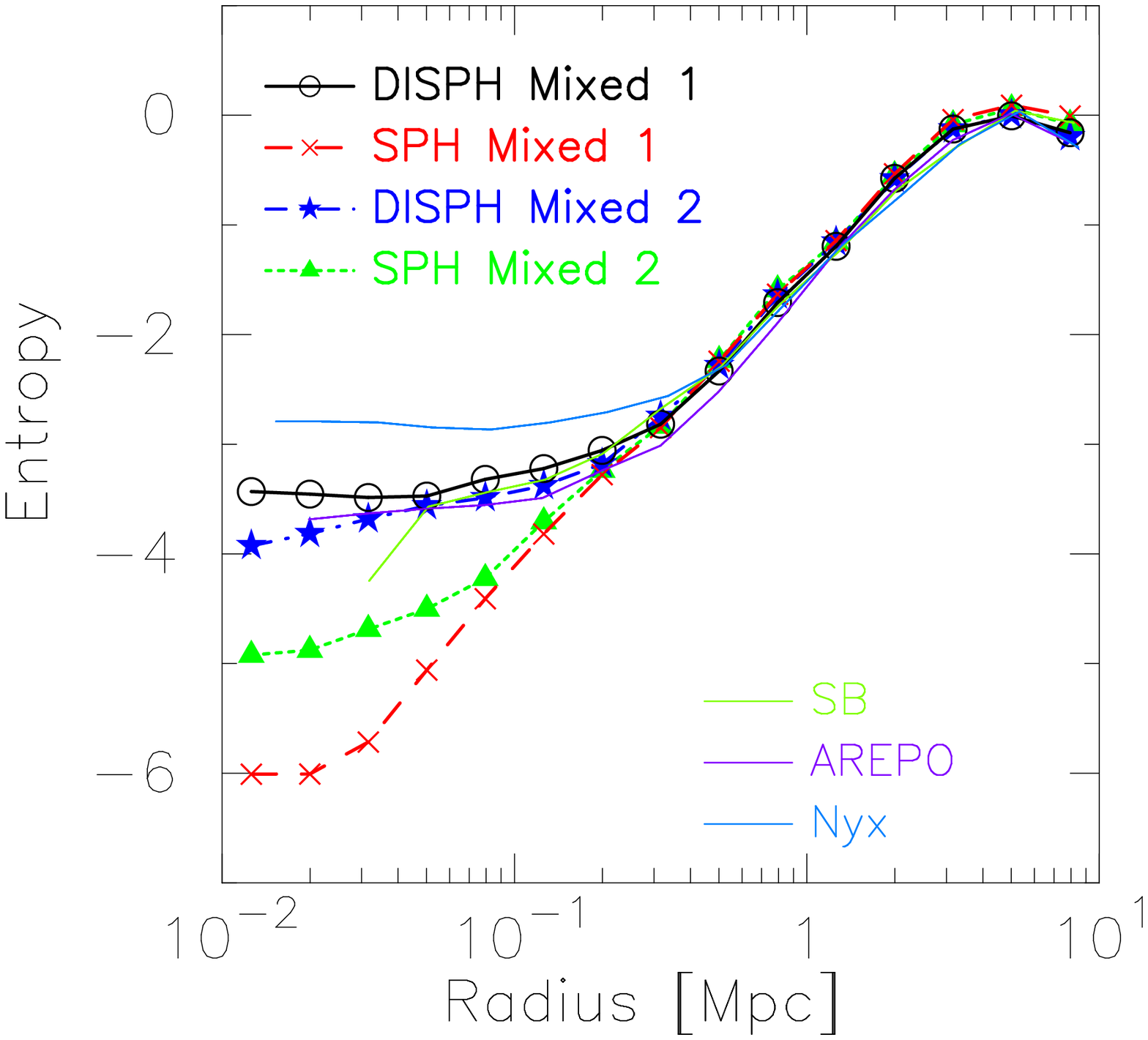}
\caption{Same as figure \ref{fig:Entropy:Comparison}, but for mixed resolutions.
The solid (black) and dashed (red) curves represent the run of the mixed
resolution 1 with DISPH and SSPH, whereas the dot-dashed (blue) and dot (green)
curves indicate the results of the mixed resolution 2 with DISPH and SSPH.
The reference curves are also shown.
\label{fig:Entropy:Mixed}
}
\end{figure}

In order to assess which parameters, i.e., mass and spatial resolutions, are more
crucial to the final entropy profile of the cluster, we carried out two extra
models with DISPH and SSPH which mix resolutions; (1) with the high mass/low
spatial resolutions (``Mixed resolution 1'') and (2) with the low mass/high
spatial resolutions (``Mixed resolution 2'').  As shown in figure
\ref{fig:Entropy:Mixed}, the former run reproduced the entropy profile of the
high-resolution SSPH run, while the latter run did that of the low-resolution
SSPH run.  A slight difference is found only in the very central region ($R_{\rm
d} < 20~{\rm kpc}$).  Hence, in the SSPH runs, the central structure of the
entropy profile is sensitive to the small scale power of the initial density
fluctuations.  There is no significant influence on the results with DISPH.

The good agreement between results with DISPH, {\tt AREPO}, and {\tt GIZMO} is
quite encouraging. However, the analysis so far is only limited at the date of
$z=0$.  In the next subsection, we compare the results obtained by DISPH and
SSPH methods deeply, by focusing on the formation process of the cluster.  

\subsection{Entropy Core/Cusp Formation} \label{sec:results:Entropy:Formation}

The cluster grew rapidly while $z>1$, and then experienced a major merger around
$z\simeq0.7$, and evolved mildly until $z=0$.  We follow these typical phases of
evolution to understand how and when the entropy core and cusp are established.

\begin{figure}
\centering
\epsscale{1.0}
\plotone{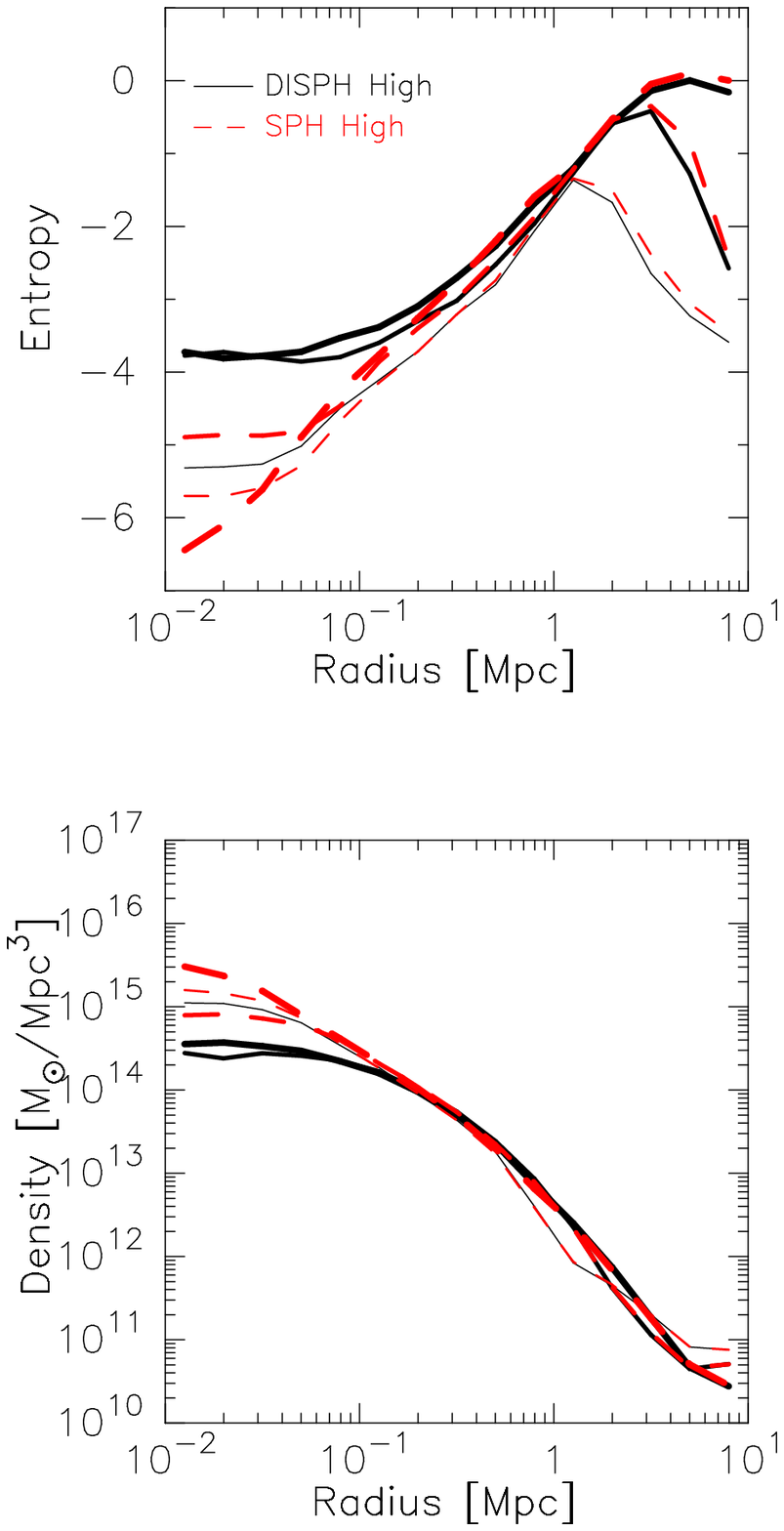}
\caption{Radial entropy (top) and density (bottom) profiles of high resolution
runs with DISPH/SSPH at three different epochs. Thin, normal, and thick curves
are the radial entropy profiles at $z=1$, $z=0.5$ and $z=0$, respectively. Solid
curves are for the high resolution DISPH run, whereas dashed ones are for the
high resolution SSPH run.
\label{fig:Entropy:ThreeEpochs}
}
\end{figure}

Figure \ref{fig:Entropy:ThreeEpochs} shows the radial entropy and density
profiles for the high-resolution DISPH and SSPH runs. We here plot them for
three epochs, before the major merger ($z=1$), after the major merger ($z=0.5$),
and at the end of the simulation ($z=0$).

For the run with DISPH, at $z=1$, the entropy profile has a relatively small core.
Then, the core entropy and core size increase and these quantities are kept
unchanged until $z=0$.  The entropy profile at $z=0.5$ is comparable to that at
$z=0$; the last major merger event triggers this core entropy increase (see
below).  The entropy of the intermediate region of $40~{\rm kpc} \le R_{\rm d}
\le 1~{\rm Mpc}$ slightly increased during $0 < z < 0.5$.  The evolution of the
density profile is similar, but in the opposite direction.  The core density
decreases by almost a factor of three from $z=1$ to $z=0.5$, and it remains
unchanged until $z=0$.

The evolution of the radial profile in the SSPH run is quite different from that
in the DISPH run.  At $z=1$, the entropy profile with SSPH has a small core
which is comparable to that of the DISPH run.  This entropy core becomes larger
until $z=0.5$.  However the core entropy is much lower than that with the DISPH run.
At $z=0$, the entropy core vanishes nearly.  The density profile follows this
evolution similarly but in the opposite direction.  At the end of the simulation
the density profile has a cusp.

\begin{figure}
\centering
\epsscale{1.0}
\plotone{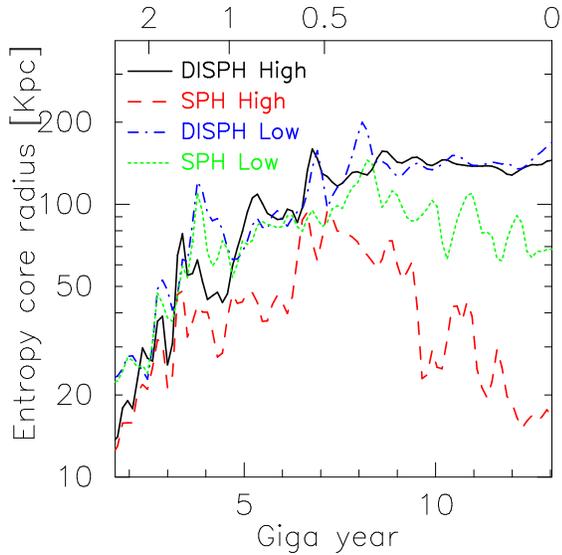}
\caption{Entropy core radius as a function of the cosmic age.  The corresponding
redshift is also shown (the digit above the panel).
Eq \ref{eq:EntropyCore} is used in order to measure the core radius. 
The solid (black), dashed (red), dot-dashed (blue), and dotted (green) curves 
are the results with the high-resolution DISPH, the high-resolution SSPH, the 
low-resolution SSPH, and the low-resolution SSPH, respectively.
\label{fig:Entropy:CoreRadius}
}
\end{figure}

Evolution of the radius of the entropy core and the averaged core entropy in the
runs with DISPH and SSPH are shown in figures \ref{fig:Entropy:CoreRadius} and
\ref{fig:Entropy:CoreEntropy}. The definition of the core radius is
described in appendix \ref{sec:method:Core}. 
The averaged core entropy is measured using the entropy of gas in the core
radius.  Their evolutions can be divided into two phases and the transition
takes place after the major merger epoch ($z\sim0.5$).

In the DISPH runs, two results agree with each other very well.  In the early phase 
($z\geq 0.5$), the radius of the entropy core increases from $\sim 20~{\rm kpc}$
($z\sim 3$) to $100~{\rm kpc}$ ($z\sim0.5$) and the averaged core entropy
increases from $-7$ ($z\sim 3$) to $\sim -4$ ($z\sim0.5$).  Then the growth of
the core radius is stopped and the core is essentially unchanged until $z=0$.

In the SSPH runs, the evolution is completely different from those of DISPH
runs.  When we compare the high resolution runs, the core radius of the SSPH run
is smaller than that of the DISPH run in the initial phase ($z\ge0.5$).  Major
merger triggered the rapid increase, but at $z\sim0.5$ the core is still smaller
than that of DISPH runs.  Second, the core size shrinks and the averaged core
entropy decreases in the late phase ($z\le0.5$).  These figures suggest that the
low-entropy material is supplied to the central region, most likely via minor
mergers in the late phase.  The low resolution SSPH run follows the evolution of
the high resolution one, although its goes between DISPH runs and the high
resolution SSPH runs because of lower resolution.

\begin{figure}
\centering
\epsscale{1.0}
\plotone{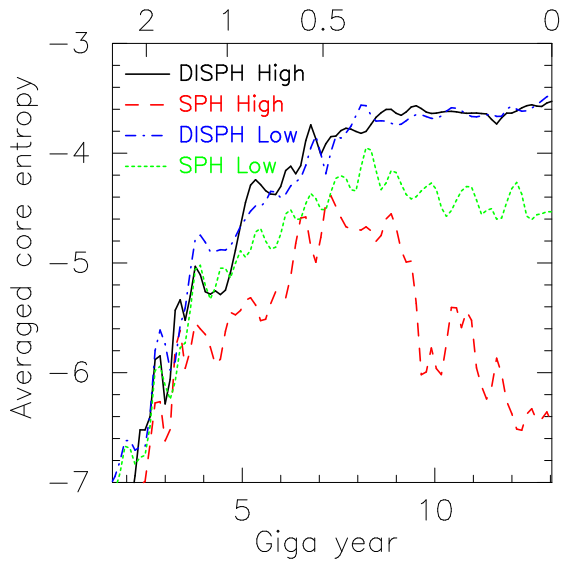}
\caption{Averaged core entropy as a function of the cosmic age.  The averaged
core entropy is calculated using particles in the core radius.  The
corresponding redshift is also shown above the plot panel.  Types of curves are
the same as figure \ref{fig:Entropy:CoreRadius}.
\label{fig:Entropy:CoreEntropy}
}
\end{figure}

In \S \ref{sec:MajorMerger} and \S \ref{sec:MinorMerger}, we further investigate
the evolution of the cluster's central entropy induced by the major and minor
mergers.

\subsubsection{Change of Core Entropy by Major Merger} \label{sec:MajorMerger}

\begin{figure*}[htb]
\centering
\epsscale{1.0}
\plotone{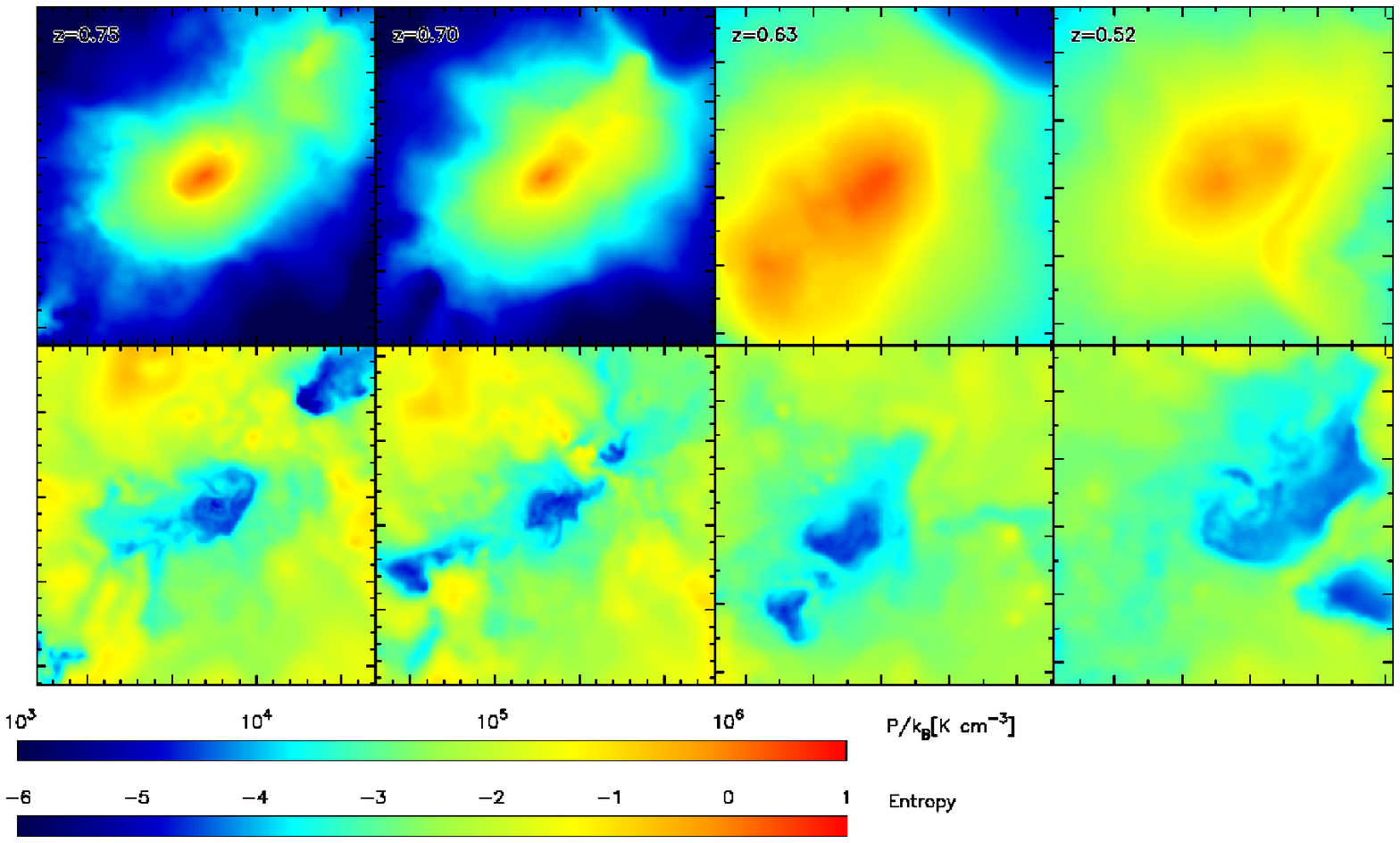}
\caption{Evolution of the cluster for the high resolution DISPH run during the
major merger phase.  Top and bottom panels show the distributions of pressure
and entropy at the $XY$ plane which across the density center evaluated by Eq
\eqref{eq:DensityCenter}.
The plot regions of the left two columns are $4~{\rm Mpc} \times 4~{\rm Mpc}$
whereas those of the right two columns are $2~{\rm Mpc} \times 2~{\rm Mpc}$.
The epoch (redshift) is displayed on the top-left corner of each panel of
pressure map.
\label{fig:Entropy:Merger:DISPH}
}
\end{figure*}

Figure \ref{fig:Entropy:Merger:DISPH} shows the evolution of the distributions
of pressure and entropy in the cluster in the high resolution DISPH run during
the major merger phase ($0.75>z>0.52$).  A smaller cluster consisting of the
low-entropy material approaches from the top-right corner ($z=0.75$), it merges
with the main cluster ($z=0.70$) and after several crossing times these two
clusters completely merge to form a single cluster with the relaxed structure
($z=0.63$ and $z=0.52$).  In this run, the entropy at the cluster center
increases due to shocks (see bottom panels). Before the merging event, the
central entropy is lower than $-4$, but it becomes higher than $-4$ after the
merger event (See also figure \ref{fig:Entropy:CoreEntropy}, $s\sim-4.4$ at
$z\sim0.74$ whereas $s\sim-4$ at $z\sim0.5$).  The pressure maps are smooth
throughout this event (Top panels).

\begin{figure*}[htb]
\centering
\epsscale{1.0}
\plotone{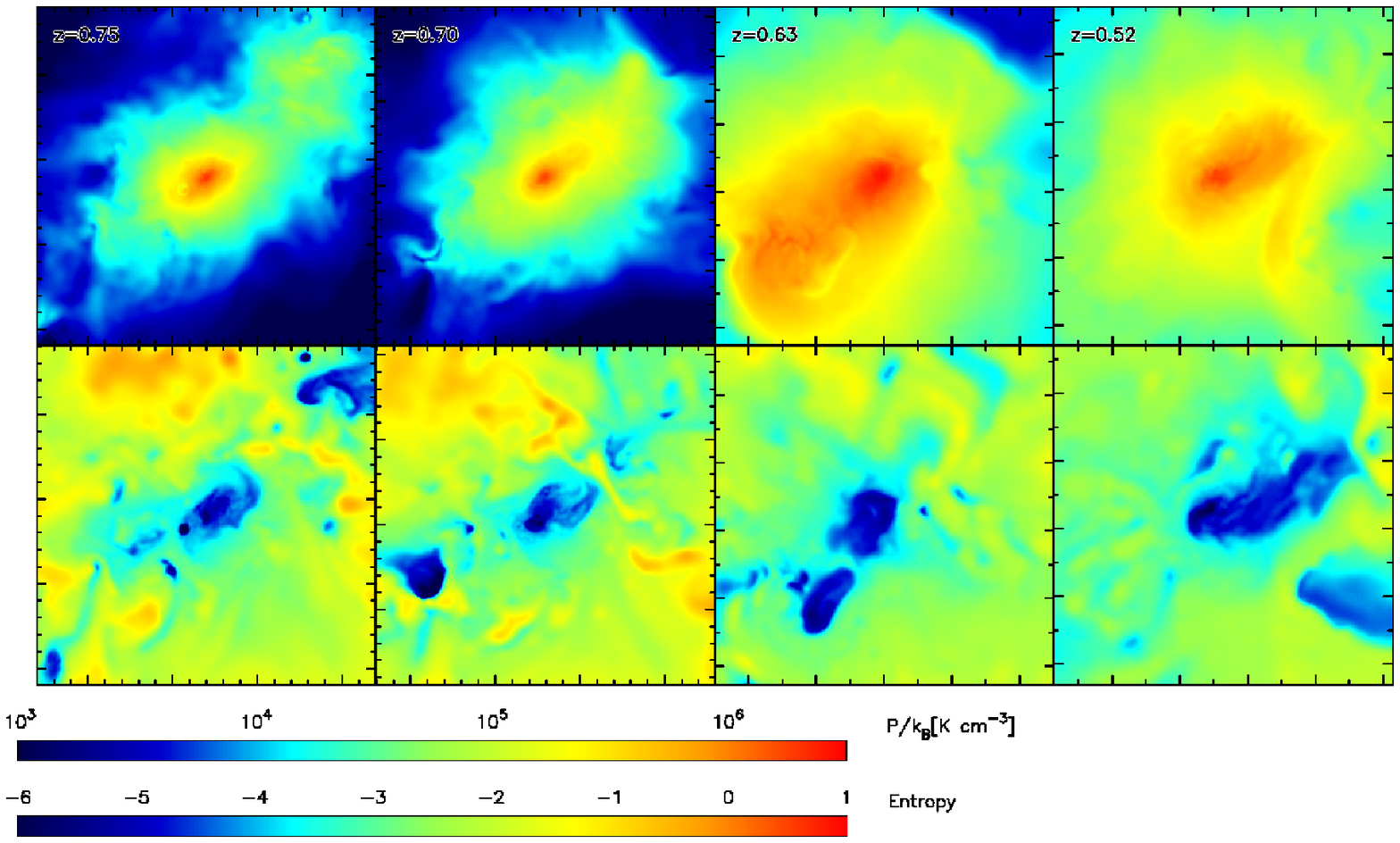}
\caption{Same as figure \ref{fig:Entropy:Merger:DISPH}, but for the
high-resolution SSPH run.
\label{fig:Entropy:Merger:SSPH}
}
\end{figure*}

In the SSPH run, we also observe the increase of the central entropy whereas the
absolute value is lower than that in the DISPH run, as we see figure
\ref{fig:Entropy:Merger:SSPH}.  There are sharp edges in the distribution of the
low-entropy (cold) gas clumps, and therefore the mixing of low- and high-entropy
components is much weaker in the SSPH run compared to that in the DISPH run.
Pressure maps show sharp gaps which are not found in those in the DISPH run.
Thus, these gaps are outcomes of the unphysical surface tension of SSPH,
resulting in the suppression of the entropy generation due to shocks.

\subsubsection{Change/Non-change of Core Entropy by Minor Mergers} \label{sec:MinorMerger}

\begin{figure*}[htb]
\centering
\epsscale{1.0}
\plotone{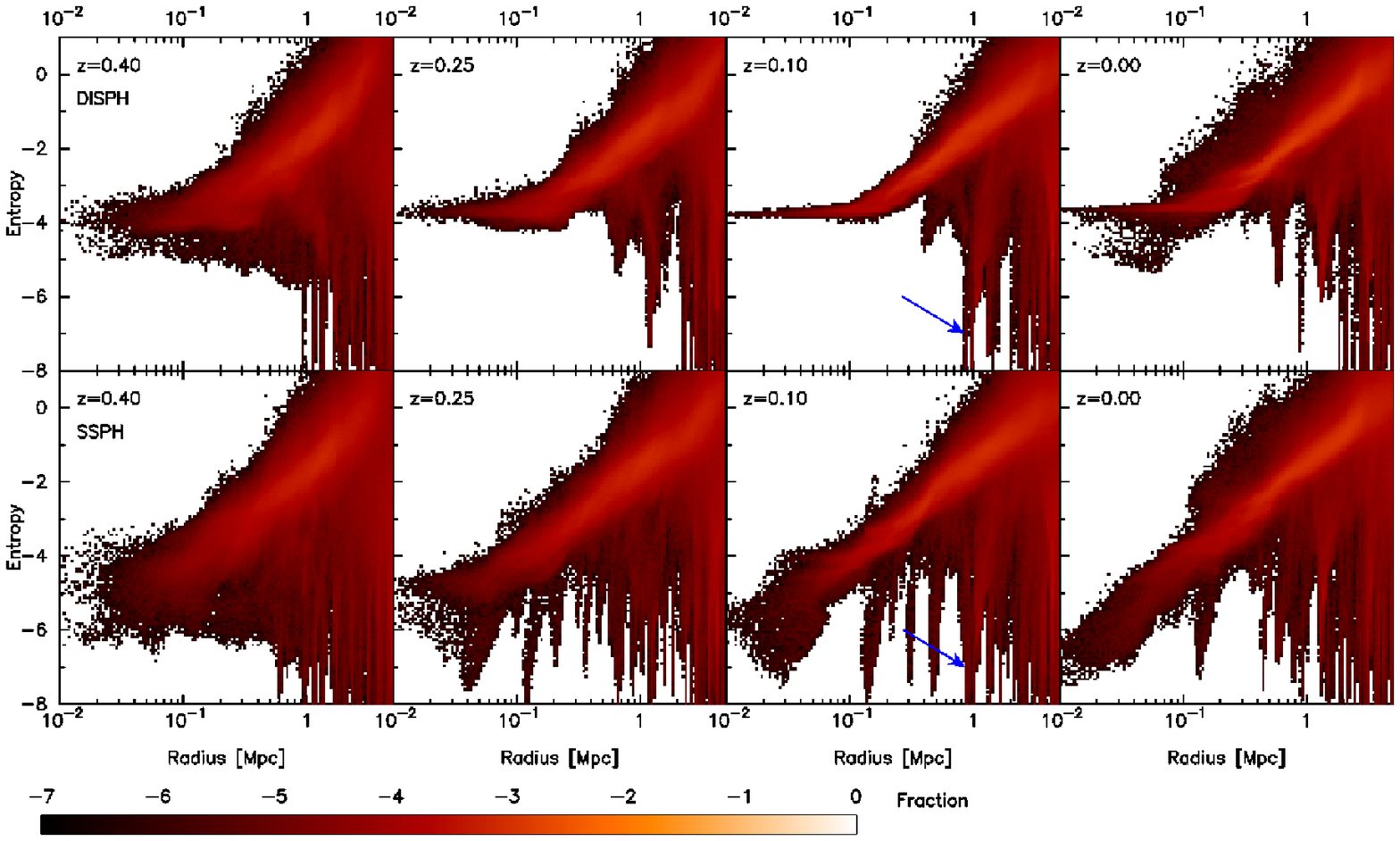}
\caption{Evolution of radial entropy distributions.  Top panels show the results
of the high resolution DISPH run, whereas bottom panels represent those of the
high resolution SSPH run. Colors correspond to the relative fraction of
particles in the logarithmic scale. Blue arrows on the panels of $z=0.1$ point
the positions of the representative clumps (see text).
\label{fig:Entropy:Merger:Minor}
}
\end{figure*}

Here, we focus on the late stage of the evolution of the cluster. 
Figure \ref{fig:Entropy:Merger:Minor} shows the evolution of the radial entropy
distribution.  The main sequence of the entropy distribution, where the relative
fraction is high, corresponds to the radial entropy profile we saw in, e.g.,
figure \ref{fig:Entropy:Comparison}.  In addition to the main sequence, there
are a number of low entropy components.  These low-entropy components are clumps
in and around the cluster.

The evolution of the cold gas clumps are quite different in the DISPH run and
SSPH run.  In the case of SSPH run (bottom panels), there are many clumps with
minimum entropy around $-8$, within the distance range $0.1~{\rm Mpc}-1~{\rm
Mpc}$.  On the other hand, in the case of the DISPH run, there are only a few
clumps within the distance $1~{\rm Mpc}$, and their minimum entropy is much
higher.  Thus, in the case of the SSPH runs, cold gas clumps survive in the
cluster and fall to the cluster center, forming low-entropy core, while in the
case of the DISPH run, cold gas clumps are disrupted before they reach to the
center of the cluster.

In order to clarify the difference of the late stage evolution of the DISPH
run and that of the SSPH run, we here focus on the evolution of one clump.
For this analysis, we first pick up low entropy component
within $R_{200}$.  The threshold entropy is $-4.5$ which is slightly lower than
the core entropy of the DISPH runs at $z=0$. With this threshold, we found a
number of low entropy clumps. In order to select one representative clump from
them, we imposed the following conditions: (1) the clump is approaching to the
cluster center, (2) the mass is sufficiently large so that we can track its
evolution, (3) the clump reaches to the center of the cluster before $z=0$, and
(4) the clump can be found in both runs at the same position.  The
representative clump we picked up is found at $R_{\rm d} \sim 1.7~{\rm Mpc}$ and
its total mass is $\sim 10^{13}~{\rm M_{\odot}}$ at $z=0.12$.

\begin{figure*}[htb]
\centering
\epsscale{1.0}
\plotone{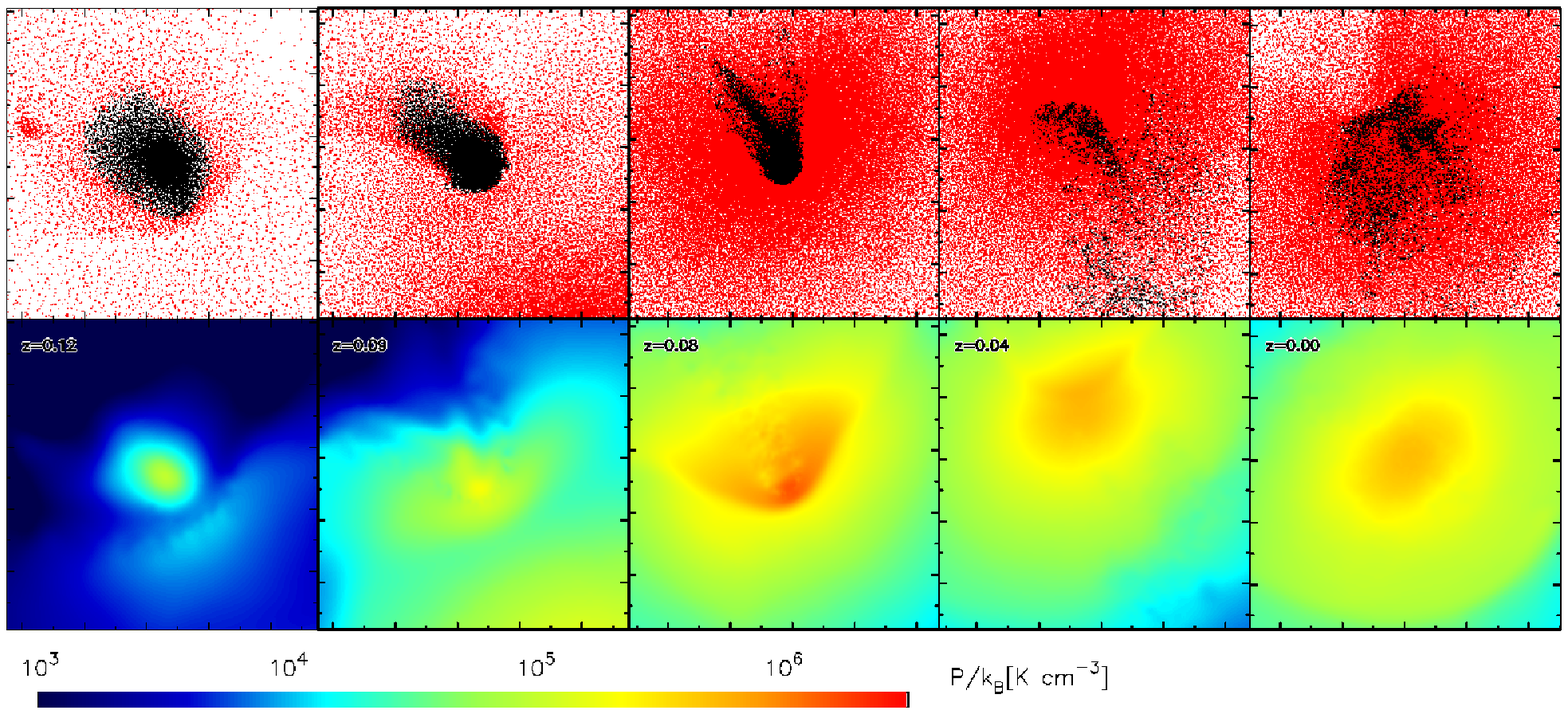}
\caption{Evolution of the low entropy clump in the high-resolution run with
DISPH.  Top panels show the distributions of particles initially associated with
the low-entropy clump (black) and the others (red).  The coordinates center is
set to the baricenter of the clump and the regions of $1~{\rm Mpc} \times 1~{\rm
Mpc}$ are plotted.  The articles shown in the upper panels are selected from
particles in the thickness of $Z = \pm 0.1~{\rm Mpc}$.  Bottom panels display
the pressure maps at the $Z=0$ plane.  The time from the left to right panels is
$\sim 2~{\rm Gyr}$.
\label{fig:Target:DISPH}
}
\end{figure*}

Figure \ref{fig:Target:DISPH} shows the evolution of this
low-entropy clump in the high-resolution DISPH run. From the particle
distribution maps (upper panels), we can see that this representative clump
enters the cluster center forming a bow-shock. Its front surface is being
stripped due to fluid instabilities, resulting in the complete destruction of
the clump by $z=0.04$ \citep[recall the blob test in][]{Agertz+2007}.  The
pressure distribution across the interface between the clump and the
intracluster medium is always smooth.

\begin{figure*}[htb]
\centering
\epsscale{1.0}
\plotone{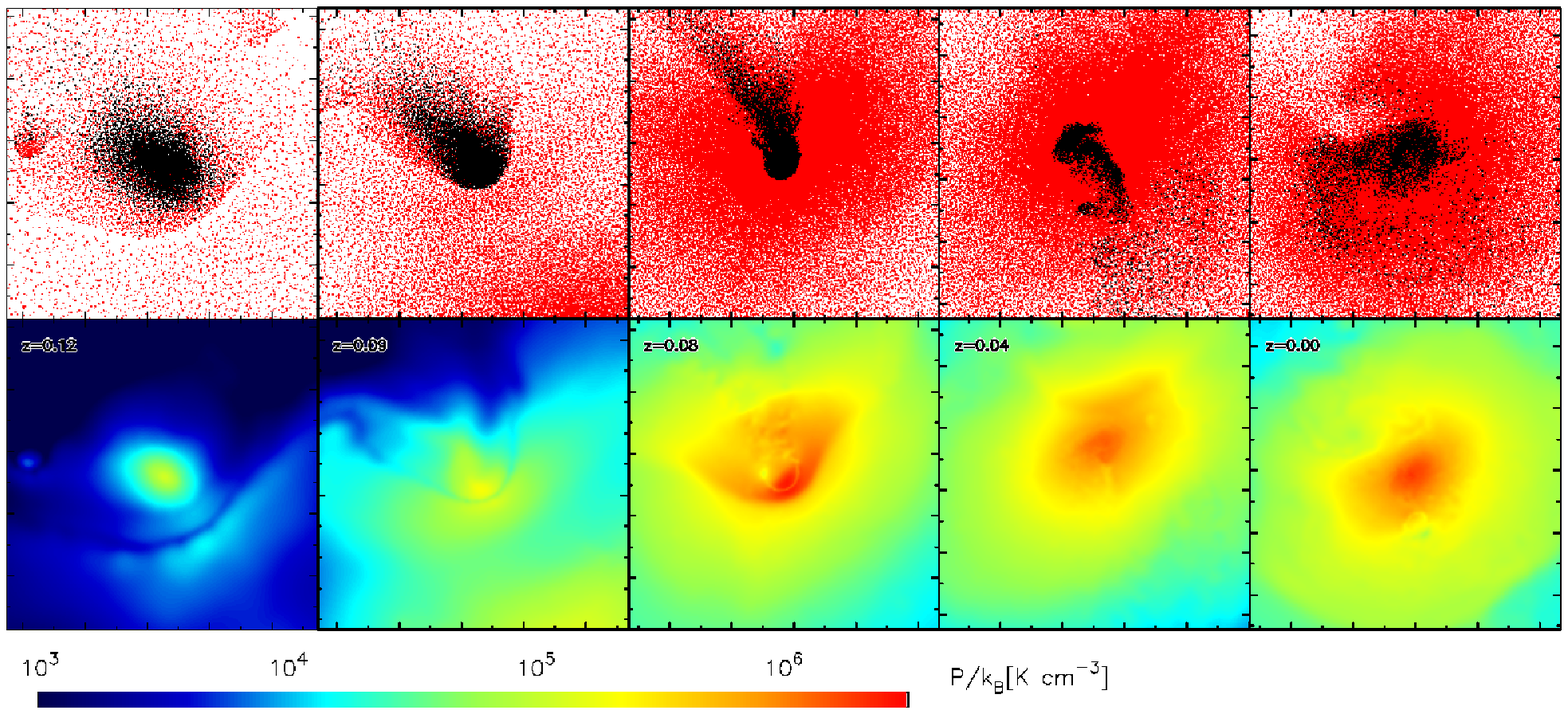}
\caption{The same as figure \ref{fig:Target:DISPH}, but for the high-resolution SSPH.
\label{fig:Target:SSPH}
}
\end{figure*}

Figure \ref{fig:Target:SSPH} shows the evolution of the same clump in the
high-resolution SSPH run.  There is a clear gap both in the particle and
pressure distributions in the front of the clump. This gap is induced by the
unphysical surface tension, protecting this clump from the development of fluid
instabilities at the surface and ram-pressure stripping.  Thus, the low entropy
gas in this clump can sink to the cluster center.

\begin{figure}[htb]
\centering
\epsscale{1.0}
\plotone{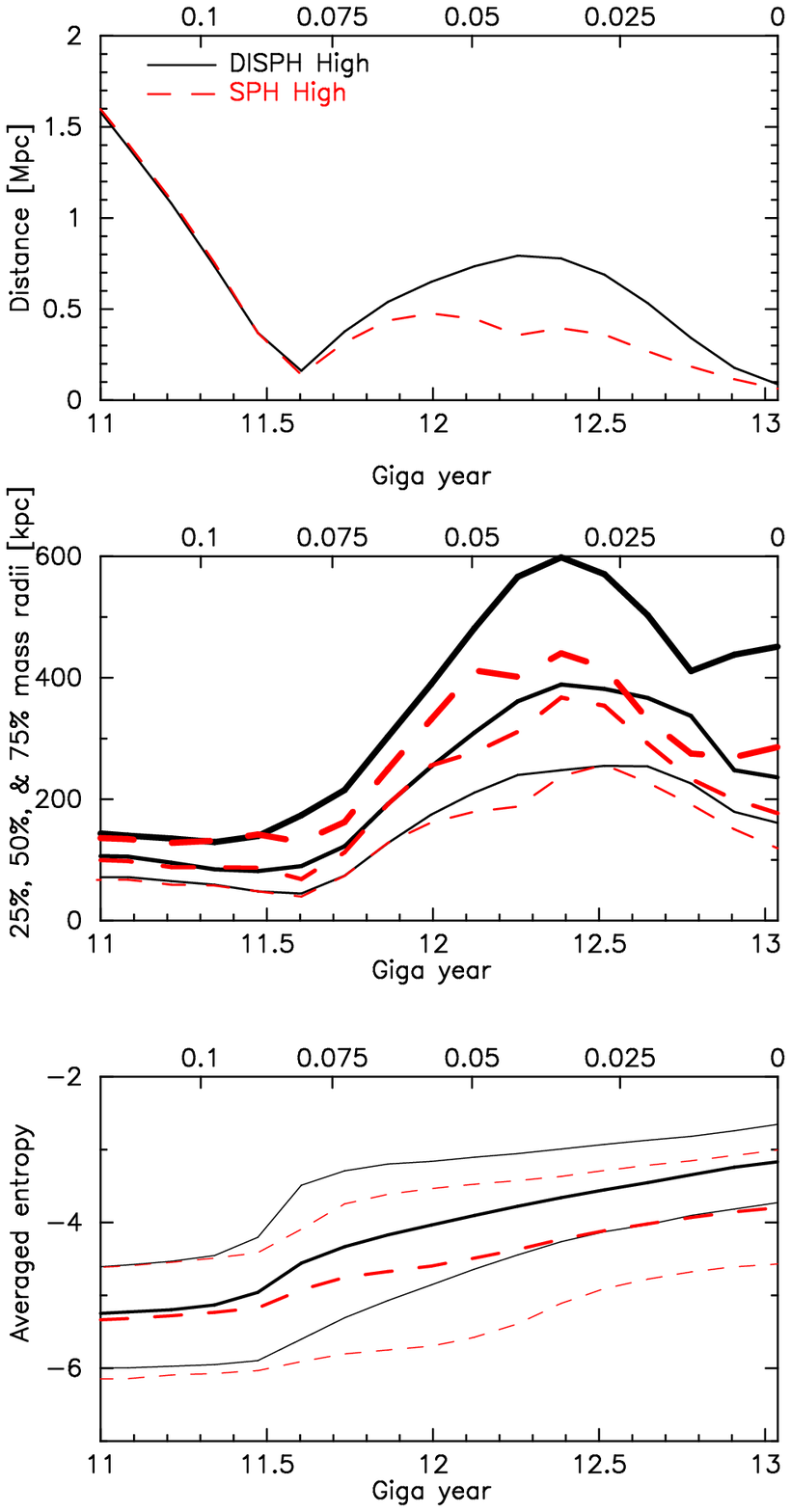}
\caption{
Time evolutions of the distance from the center of the cluster, the size and the
mean entropy of the representative clump.  Top: Distance between the entropy
weighted center of the representative clumps and the density peak of the halo.
Middle: $25\%$ (thin curve), $50\%$ (normal curve) and $75\%$ (thick curve)
radii from the clump center.  Bottom: Mean entropies of the gas in the
representative clumps (thick curves).  Thin curves for each run represent $10\%$
and $90\%$ spread of the entropy distribution.
\label{fig:Target:Evolutions}
}
\end{figure}

We compare the evolutions of this representative clump in DISPH and SSPH runs in
figure \ref{fig:Target:Evolutions}.  Top, middle, and bottom panels are the
distance from the cluster center, the size, and the averaged entropy of the gas
particles of the clump, respectively.  Before the pericenter passage, clumps in
two runs are almost identical.  When it passed the pericenter, the orbital decay
starts. The orbital decay in the SSPH run is much faster than that in the DISPH
run.  Clumps expand when passing the pericenter.  The clump in the DISPH run is
more extended compared to that in the SSPH run.  These two results tell us that
the clump in the SSPH run keeps a compact structure much longer.

The increase of the mean entropy starts slightly before the pericenter passage
in both runs.  At $z=0$, the mean entropy of the clump in DISPH is $\sim 1$
higher than that in SSPH. This indicates that the shock heating is more
efficient in the DISPH run.  However, behind the fact that SSPH has unphysical
surface tension, we should say that the shock heating in SSPH is more
inefficient. DISPH can capture the intrinsic shocks.  This difference in the
shock heating efficiency explains why cold gas clumps can reach to the
center of the cluster in the SSPH run but are disrupted in the DISPH run.

This tendency that the clumps in the SSPH run survive longer at the center of the
cluster compared to those in the DISPH run can be seen in figure
\ref{fig:Target:Stack}. Here, we used subfind clumps (see appendix
\ref{sec:method:Subfind}) and stacked data during $0 \leq z \leq 0.12$ (total 17
snapshots).  Generally, as clumps move closer to the center of the cluster, they
loose their masses due to tidal and ram-pressure stripping.
At the same distance from the cluster center, in particular for $R<1~{\rm Mpc}$,
we can see the clear tendency that clumps in SSPH are more massive than those in
DISPH.  Thus, the evolution of the inner part of the cluster strongly affected
by the choice of the SPH scheme.  While clump surviving time is different, there
is no significant difference in mass functions with different SPH schemes (see
appendix \ref{sec:results:MassFunction}).  

\begin{figure}[htb]
\centering
\epsscale{1.0}
\plotone{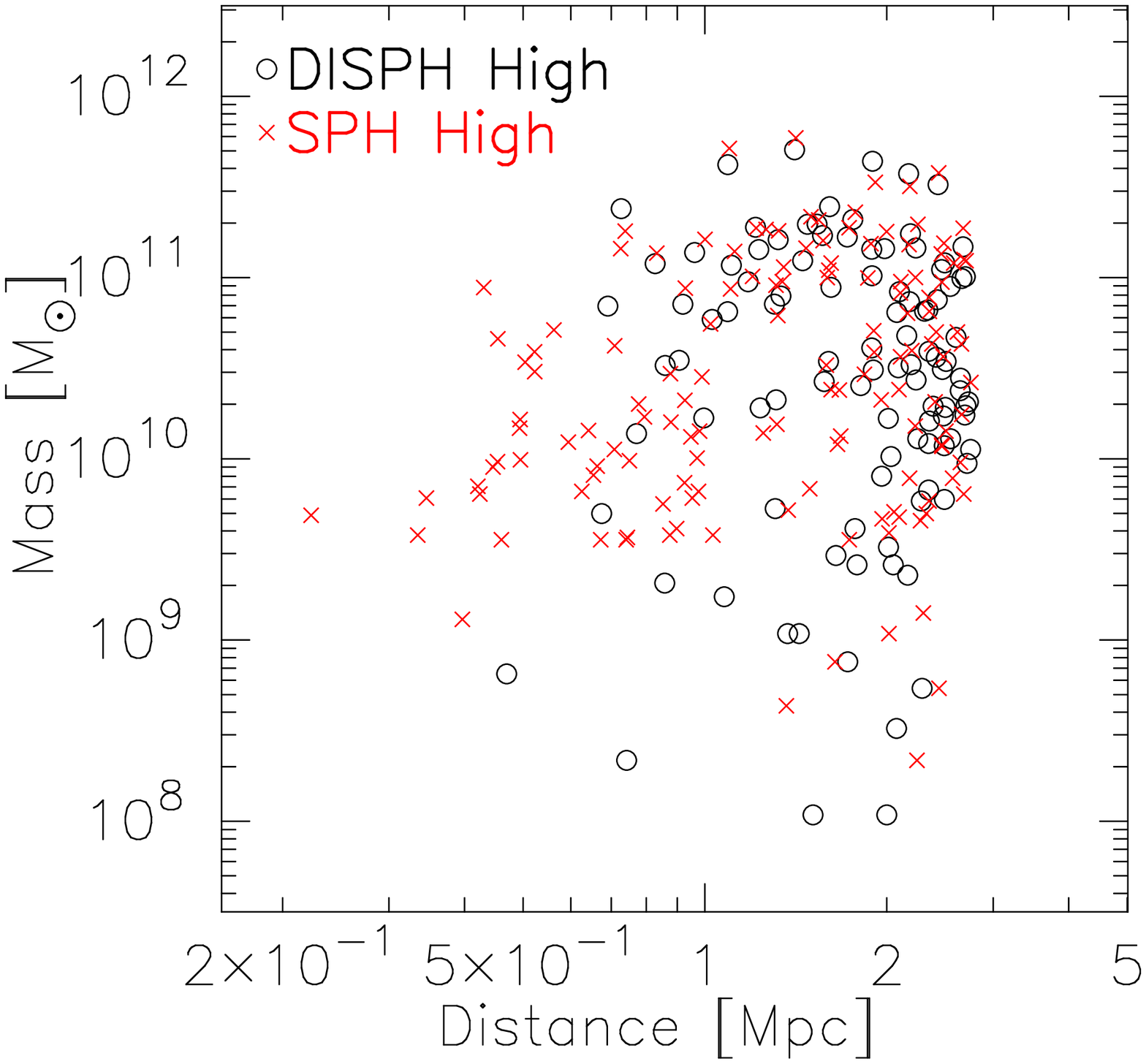}
\caption{Masses of the gas component in subfind clumps as a function of the
distance from the cluster center.  We stacked clumps within $R_{200}$ from
$z=0.12$ to $z=0$.  Total number of the snapshots is 17 and the time interval
between two closest snapshots are $\sim 130~{\rm Myr}$.  Open circles are for
the high-resolution DISPH run whereas crosses are for the high-resolution SSPH
run.  \label{fig:Target:Stack}
}
\end{figure}

It has been pointed out by \cite{Power+2014} that the cluster core entropy
decreased at the low-$z$ with SSPH.  They measured the evolution of entropy at
the central region ($0.01\times R_{200}$) of clusters of galaxies in a
$\Lambda$CDM universe. They found that the core entropy of the SSPH run
decreased continuously for $0 < z < 0.6$ whereas those of AMR and SPHS runs were
roughly constant.  Although they did not follow the individual clumps, they
explained that the change of the central entropy found in low-$z$ period is due
to the accretion of the low-entropy clumps induced by the minor mergers.  Our
study supports their result and this process works in the evolution of the Santa
Barbara cluster in previous studies as well.

\subsection{Effects of Artificial Conductivity Term} \label{sec:results:AC}

Recently, the AC term has become widely used in SPH simulations in order to remove the
unphysical surface tension \citep[e.g.,][]{Price2008, Rosswog2009,
ReadHayfield2012, Kawata+2013, Beck+2016}.  Here, we investigate the effects of
AC on the cluster formation and evolution.  Similar studies are found in the
literature \cite[e.g.,][]{Wadsley+2008, Power+2014, BiffiValdarnini2015,
Hopkins2015, Beck+2016, Sembolini+2016}.

According to the numerical experiments by \cite{Wadsley+2008} and
\cite{Power+2014}, it is expected that the entropy core appears if the AC term
is introduced into SSPH and the core size and core entropy depend on the maximum
value of the conduction control parameter $\alpha_{\rm AC,max}$.  Thus, in this
section, we highlight the contribution of $\alpha_{\rm AC,max}$ to the thermal
structure and the entropy profile.

Here, we adopt the following functional form for the AC:
\begin{equation}
\frac{du_i}{dt} = \sum_j m_j 
\frac{4 \alpha_{{\rm AC},i}\alpha_{{\rm AC},j}}{\alpha_{{\rm AC},i}+\alpha_{{\rm AC},j}}
\frac{v_{{\rm sig},ij}^{\rm u}}{\rho_{ij}} (u_i-u_j) 
\nabla W_{ij}, \label{eq:AC}
\end{equation}
where $\rho_{ij} = 0.5 (\rho_i+\rho_j)$, $v_{{\rm sig},ij}^{\rm u} =
\sqrt{|P_i-P_j|/\rho_{ij}}$, $P$ is pressure, and $\nabla W_{ij} = 0.5 [ \nabla
W(\boldsymbol x_j- \boldsymbol x_i,h_i)+ \nabla W(\boldsymbol x_i- \boldsymbol
x_j,h_j)]$. 
Following \cite{Beck+2016}, we evaluate the AC coefficient of particle $i$,
$\alpha_{{\rm AC},i}$ using the following equation;
\begin{equation}
\alpha_{{\rm AC},i} = \frac{h_i}{3} \frac{|\nabla u_i|}{|u_i|},
\end{equation}
where $\nabla u_i$ is evaluated by the SPH manner and
$\alpha_{{\rm AC},i}$ is allowed to evolve from $10^{-5}$ to $\alpha_{\rm AC,max}$.
This AC term is essentially the same as those adopted in the previous studies
\citep[e.g.,][]{Price2008, Rosswog2009, ReadHayfield2012, Power+2014, Hu+2014,
Hopkins2015, Beck+2016}. We tested several functional forms and switches
and found that the conclusion in this subsection does not affected by them. 
We show five high-resolution SSPH runs with different values of $\alpha_{\rm
AC,max}$: $\alpha_{\rm AC,max} = 0.01, 0.1, 0.25, 1$ and $5$.

\begin{figure*}[htb]
\centering
\epsscale{1.0}
\plotone{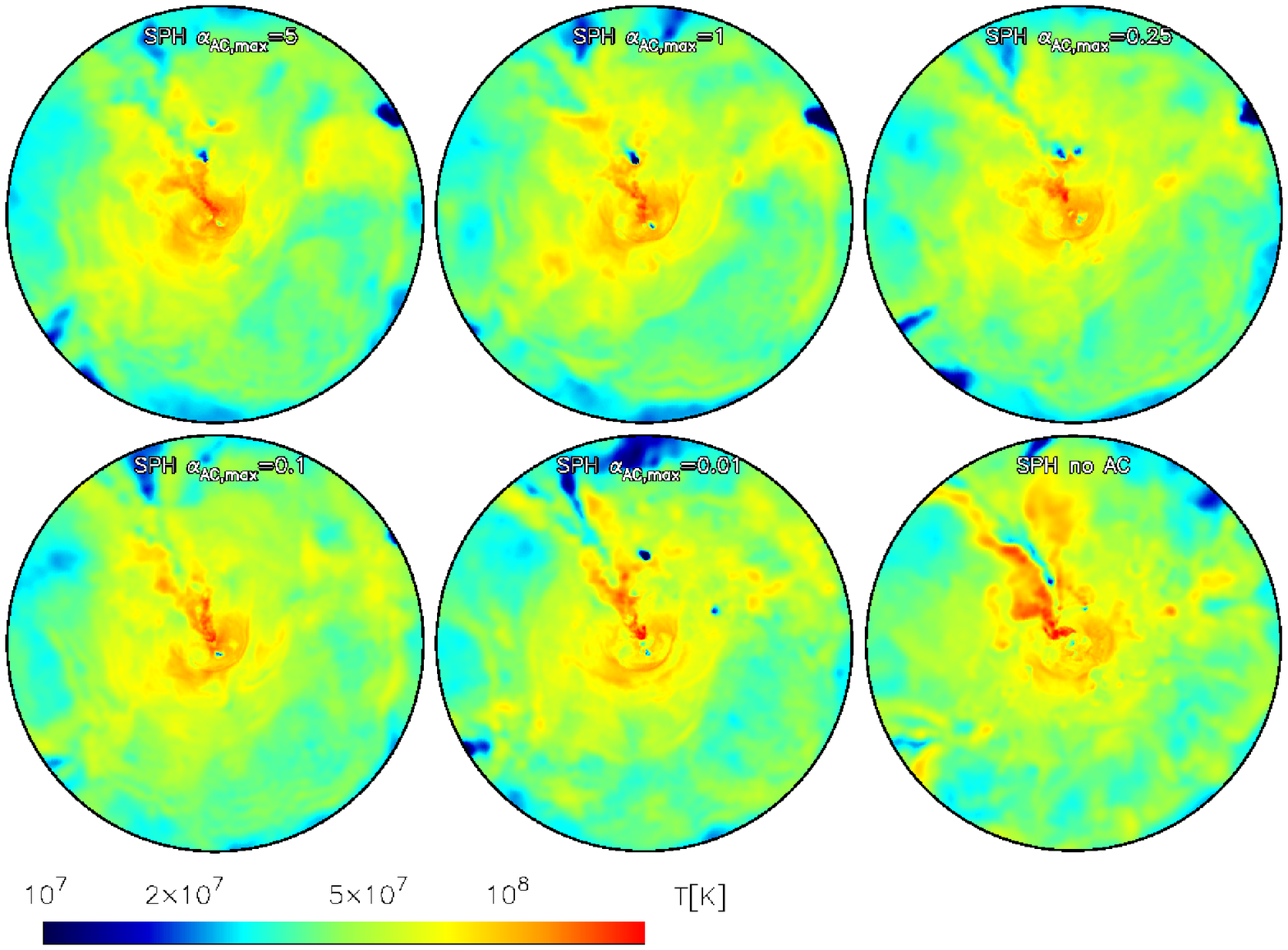}
\caption{Temperature maps of SSPH runs with the AC term at
$z=0$. Results with five different values of $\alpha_{\rm {AC,max}}$ are
displayed. The result without AC term is also shown as a
reference. The plotted region corresponds to the virial radius.
}\label{fig:AC:Temp}
\end{figure*}

Figure \ref{fig:AC:Temp} shows the temperature maps of five runs with the AC
term, as well as that of the run without AC.  From this figure, we can see that
the temperature structure gets blurred when $\alpha_{\rm AC,max}$ increases. We
find that there is a similar effect in both the density and entropy maps.  This
blurring effect has an influence on the entropy profile of the cluster, as we
see next.

\begin{figure}[htb]
\centering
\epsscale{1.0}
\plotone{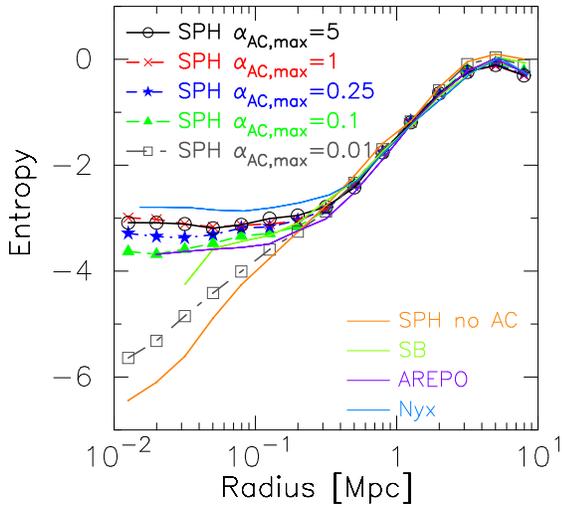}
\caption{Radial entropy profiles of SSPH runs with the AC term at $z=0$. Those
without the AC term and obtained from literatures are also shown.
}\label{fig:AC:Entropy}
\end{figure}

Figure \ref{fig:AC:Entropy} shows the radial entropy profiles with and without
AC.  Apparently, the entropy profile is affected by AC.  The entropy profile
with $\alpha_{\rm {AC,max}} = 0.01$ is cuspy; it has a slightly higher entropy
than that without AC in $R<200~{\rm kpc}$.  On the other hand, those with
$\alpha_{\rm {AC,max}} \ge 0.1$ have cores. These core entropies increase with
increasing $\alpha_{\rm {AC,max}}$ and the increase stalls when $\alpha_{\rm
{AC,max}} \ge 1$.  The core entropy with $\alpha_{\rm {AC,max}} = 0.1$ is
comparable to those with DISPH and {\tt AREPO} whereas that with $\alpha_{\rm
{AC,max}} = 5$ is rather close to that obtained by a mesh code.  In our test,
unlike previous study \citep{Beck+2016}, the core entropies with $\alpha_{\rm
{AC,max}} \ge 1$ are slightly lower than those obtained by mesh codes.  This
difference might come from the detailed implementations.

\begin{figure}[htb]
\centering
\epsscale{1.0}
\plotone{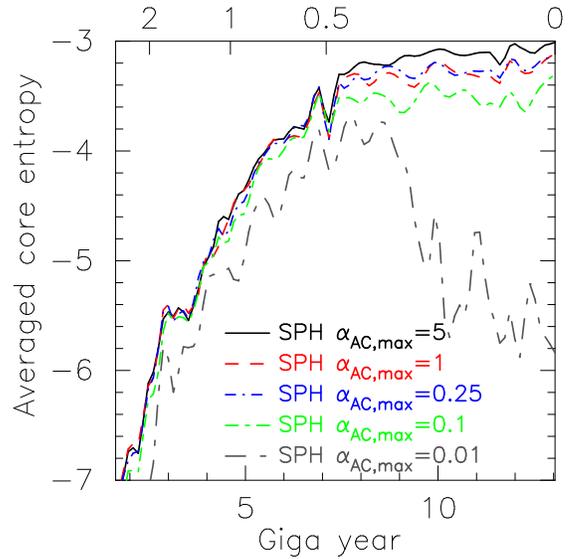}
\caption{
Averaged core entropy as a function of the cosmic age. Results of five AC runs
are shown.
}\label{fig:AC:Entropy:Core}
\end{figure}

Time evolutions of averaged core entropies for five AC runs are shown in figure
\ref{fig:AC:Entropy:Core}.  The overall evolutions of the runs with $\alpha_{\rm
AC,max} \ge 0.1$ are similar to those of DISPH runs; the increase of the core
entropy is fast until $z\sim0.5$ and then it stops.  On the other hand, the
evolution with $\alpha_{\rm AC,max} = 0.01$ is close to that without AC, whereas
this run has a slightly higher entropy within $200~{\rm kpc}$.  The final core
sizes are $220$--$280~{\rm kpc}$ for the runs with $\alpha_{\rm AC,max} \ge 0.1$
(Recall that they are $\sim 150~{\rm kpc}$ in DISPH runs). 

Although SSPH runs with AC ($\alpha_{\rm AC,max} \ge 0.1$) have slightly larger
entropy cores, the effect of the AC term looks similar to that of DISPH; the
essential difference is that AC requires the new free parameter $\alpha_{\rm
AC,max}$. Since the numerical results depend on this parameter, one needs to pay
attention to determine its value.  It is of course possible that the calibrated
$\alpha_{\rm AC,max}$ for one experiment is inadequate for another experiment.

\section{Summary \& Discussion} \label{sec:discussion}

In this paper, we studied the formation of the entropy core in the Santa Barbara
cluster.  Previous studies \citep[e.g.,][]{Frenk+1999,
Kravtsov+2002, Springel2005, Almgren+2013} have demonstrated that simulations
with Eulerian mesh codes make the entropy core in the cluster center while those
with Lagrangian SPH codes do not.  The original paper \citep{Frenk+1999}
argued that the origin of this difference came from the different treatment of
shocks in the SPH and mesh codes.  However, both Eulerian mesh codes and
Lagrangian SPH codes have their own, intrinsic limitations which might affect
the entropy evolution.  This implies that neither mesh results nor SPH results
is correct.  \cite{Springel2010} and \cite{Hopkins2015} reported that new
Lagrangian fluid schemes made entropy cores but the depth of their cores is
deeper than those obtained by Eulerian codes.

The clusters simulated with DISPH formed prominent entropy cores and its size
and absolute value of the core entropy was insensitive to the adopted
resolutions.  During the cluster formation, a number of low entropy clumps
accreted and these clumps brought low entropy materials to the cluster center.
In the DISPH runs, the ram pressure stripping and fluid instabilities disrupted
these clumps efficiently in DISPH runs.  The entropy of these clumps increased
due to hydrodynamical shock and they built the entropy core.

The size of the entropy core at $z=0$ is $\sim 150~{\rm kpc}$ and the depth of
that is $s \sim -4$. These values are comparable with those obtained by the
original version of the moving mesh code \citep{Springel2010} and the mesh free code
\citep{Hopkins2015}. The values of the core entropy obtained in these simulations
are about unity lower than those with Eulerian mesh codes.  This systematic
difference is probably due to the numerical diffusion intrinsically existing in
Eulerian codes \citep[e.g.,][]{Tasker+2008, Springel2010}. In
principle, the effect of the numerical diffusion can be reduced by increasing
the spatial resolution \citep{Robertson+2010}.  However, whether this strategy
is really useful for cosmological simulations or not is unclear since this
approach is quite time-consuming.

We were able to reproduce the cuspy entropy profile when we used SSPH and the
results with SSPH depended strongly on the adopted mass resolution.  When we
employ low mass resolution, there is a small core which is comparable with the
previous simulations. When we adopt the high mass resolution, the central
entropy became much lower and the cusply profile appeared.  By following the
evolution of the central entropy, we found that the central entropy in SSPH runs
affected by the minor mergers involving the low-entropy gas.  Low-entropy gas in
simulations with SSPH is ``protected'' against ram-pressure stripping, as found
by \cite{Agertz+2007}.  The evolution of low-entropy clumps in DISPH is
completely different from that in SSPH.  The cuspy entropy profile found in the
SSPH runs is artifact.

We can obtain the entropy core in the Santa Barbara cluster with just changing
the volume element of the formulation and without introducing any diffusions.
There are several studies in which entropy core was obtained by introducing
dissipations to SPH \citep{Wadsley+2008, ReadHayfield2012, BiffiValdarnini2015,
Beck+2016}.  In their studies, they could obtain cored entropy profiles.
We also confirmed it by our numerical experiments.  As is reported
by previous studies, the core size depended on the diffusion coefficient of the
conductivity.  Our method has advantages over these methods since there is no
need to calibrate the diffusion coefficients.  In addition, there is no time and
spacial scales which controls the degree of the dissipation strength if we used
DISPH, since formulation of DISPH itself can remove the unphysical surface
tension.  We thus conclude that DISPH is a good alternative to SSPH for the
cosmological structure formation.

There is room to improve the implementation of the AV term and switches.  For
instance, a sophisticated switch is important in order to deal with a rotating
system \citep{CullenDehnen2010}.  Recently, \cite{Hosono+2016} showed a survey
of the implementation of the AV.  In their study, they showed that the run with
the combination of the von-Neumann-Richtmyer-Landshoff type AV term
velocity gradient \citep{garcia-Senz+2012, Hu+2014, Rosswog2015} can maintain
the Keplerian disk for $\sim 100$ orbits.  We need to investigate the
feasibility of these methods in structure formation simulations.

\acknowledgements

We thank the anonymous referee. Her/his insightful comments improve our
manuscript greatly.
T.R.S. thanks Takashi Okamoto and Kohji Yoshikawa for useful discussion.  T.R.S.
also thanks Lucy Kwok and Junko Saitoh who helped in writing the manuscript.
Numerical simulations were carried out on the Cray XC30 system in the Center for
Computational Astrophysics at the National Astronomical Observatory of Japan.
This work is supported by Grant-in-Aid for Scientific Research (26707007) of
Japan Society for the Promotion of Science and Strategic Programs for Innovative
Research of the Ministry of Education, Culture, Sports, Science and Technology
(SPIRE).  This work is supported in part by MEXT SPIRE and JICFuS.

\begin{appendix}

\section{Difficulties of AC term for a fluid consisting of
different chemical compositions} \label{sec:AC}

The philosophy of the introduction of AC is to ensure all physical quantities
will be smooth everywhere.  At a contact discontinuity, SSPH makes a pressure blip if
the internal energy does not have a smoothed profile consistent with the smoothed
profile of density. To guarantee the consistent profile of the internal
energy, AC spreads the internal energy until the pressure profile becomes smooth.
This term works very well in the standard tests and widely used.  However, one
meets a difficulty when one applies AC to a contact discontinuity consisting
of a fluid with different chemical compositions.

Consider the case in which there are two contacting fluids and they have different
chemical compositions. Here, we assume that they have different mean molecular
weights, $\mu$s. Thus, the internal energy is a function of $\mu$ and
temperature $T$; $u=u(\mu,T)$. 
Moreover, these two fluids are initially in the pressure and temperature equilibrium state.
DISPH can reproduce the original pressure and temperature equilibrium state
without any difficulty.  On the other hand, SSPH cannot reproduce it; a
pressure jump appears across the contact interface.  SSPH with the AC term gives
a pressure equilibrium state by changing the internal energy of particles around
the contact interface.  

Since $u=u(\mu,T)$, this change of $u$ induced by AC is regarded as the changes
of $T$ and $\mu$.  Unfortunately, we do not know an appropriate procedure to
spread $T$ and $\mu$.  Hence, we cannot predict exact quantities of $\mu$ and
$T$ at the contact interface.  There are two possible limits: (1) $T$ changes
but $\mu$ does not, and (2) $\mu$ alters but $T$  does not.  The former leads to
undesirable chemical reactions if one is solving the chemical reactions, since
they depend on $T$.  The latter is the compulsive mixing of chemical
compositions.

We show a simple, one-dimensional numerical experiment regarding this problem.
The initial condition is as follows.  The computational domain is $0\le x < 1$
with a periodic boundary condition.  The fluid at $0.25 \le x < 0.75$ has $\mu_1
= 1$ while the other has $\mu_2 = 2$.  Both pressure and temperature have
constant values in the whole region.  Since $\mu_1 \neq \mu_2$, the density and
internal energy have different values in these two sub-domains.  Here the
initial pressure and temperature are unity and $\gamma = 5/3$. The density is
normalized so that $\rho = \mu$.  We solved this system until $\sim100$ sound
crossing time. We used two schemes: one was DISPH and the other was SSPH with the
AC term.  Here, we adopt the AC term described in \S \ref{sec:results:AC} and
$\alpha_{\rm AC,max} = 1$.

\begin{figure}[htb]
\centering
\epsscale{1.0}
\plotone{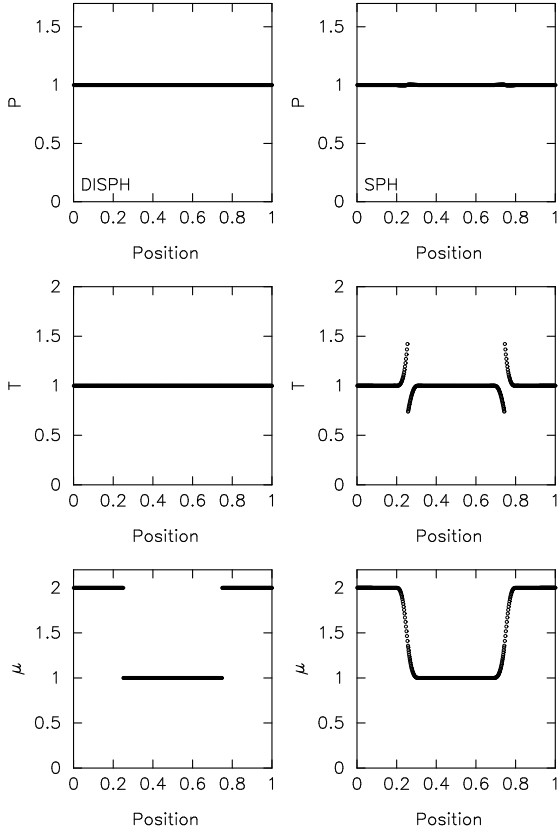}
\caption{Distributions of $P$, $T$, and $\mu$ at time 100. Left and right
columns show results of DISPH and SSPH, respectively. For SSPH, we used with the
AC term.  Top and bottom are distributions of pressure and $\mu$, respectively.
\label{fig:AC}
}
\end{figure}

Figure \ref{fig:AC} shows the results of SSPH with the AC term and DISPH at time
$100$ (about 100 crossing time).  In the SSPH run with the AC term, the initial
pressure jump is smeared by smoothing the distribution of the internal energy
(the top-right panel in figure \ref{fig:AC}).  If we consider this change of the
internal energy as the change of the temperature, we have the temperature
distribution with jumps at the contact discontinuities shown in the middle-right
panel.  On the other hand, if we regard it as the change of $\mu$, we obtain the
smoothed distribution of $\mu$ as shown in the bottom-right panel.  In DISPH,
there is no difficulty in dealing with this system (panels in the left column).

A system consisting of materials with different chemical compositions is not an
unphysical situation.  For instance, for a giant impact simulation, one solves
collisions of core(iron)-mantle(silicon based rock) planet embryos
\citep{Benz+1987, Cameron1997, CanupAsphaug2001, Canup2004,
NakajimaStevenson2014, Hosono+2016GI}.  At the core-mantle boundary, there is a
jump of the composition.  In addition, at the contact interface at the colliding
time there is also a jump of the chemical composition.  If we use the AC term
for this simulation, the boundary materials are mixed and/or the temperature at
these boundaries changes.  DISPH for a non-ideal gas \citep{Hosono+2013} can
deal with these boundaries without any difficulty \citep{Hosono+2016GI}.

\section{Time integration scheme and benchmark test} \label{sec:method:TimeIntegration}

The time integration of a single step from $n$-th step to $(n+1)$-th step where
the time step is $\Delta t$ is carried out by the following procedure:
\begin{enumerate}
\item Advance the velocity $\boldsymbol v^n$ from $n$-th step to $(n+1/2)$-th step using the
acceleration $\boldsymbol a^n$:
\begin{equation}
\boldsymbol v^{n+1/2} = \boldsymbol v^{n}+\frac{1}{2} \boldsymbol a^n \Delta t.
\end{equation}
\item Drift the position $\boldsymbol x^n$ from $n$-th step to $(n+1)$-th step
using $\boldsymbol v^{n+1/2}$:
\begin{equation}
\boldsymbol x^{n+1} = \boldsymbol x^{n}+\boldsymbol v^{n+1} \Delta t.
\end{equation}
\item Evaluate the predictor of the velocity at $(n+1)$-th step using
$\boldsymbol a^n$:
\begin{equation}
\boldsymbol v^{n+1}_{\rm p} = \boldsymbol v^{n+1/2}+\frac{1}{2} \boldsymbol a^n \Delta t.
\end{equation}
\item Evaluate the predictor of the specific internal energy at 
$(n+1)$-th step using the time derivative of $u^n$, $du^{n}/dt$:
\begin{equation}
u^{n+1}_{\rm p} = u^{n}+ \left ( \frac{du}{dt} \right )^{n} \Delta t.
\end{equation}
\item Compute $\boldsymbol a^{n+1}$ and $du^n/dt$ using $\boldsymbol x^{n+1}$,
$\boldsymbol v^{n+1}_{\rm p}$, and $u^{n+1}_{\rm p}$.
\item Advance $\boldsymbol v^{n+1/2}$ by the rest of the half step using
$\boldsymbol a^{n+1}$:
\begin{equation}
\boldsymbol v^{n+1} = \boldsymbol v^{n+1/2}+\frac{1}{2} \boldsymbol a^{n+1} \Delta t.
\end{equation}
\item Compute $u^{n+1}$ using both $du^{n}/dt$ and $du^{n+1}/dt$:
\begin{equation}
u^{n+1} = u^{n}+ \frac{1}{2} \left [ \left ( \frac{du}{dt} \right )^{n} + \left ( \frac{du}{dt} \right )^{n+1} \right ] \Delta t.
\end{equation}
\end{enumerate}

The manner to integrate the internal energy described above is Heun's method (or
the modified Euler's method or the second order Runge-Kutta method).  Thus, our
integration scheme is categorized as the second order method.  For DM particles,
this time integration scheme is reduced the standard second order symplectic
scheme, the leap-frog scheme. 

As a demonstration, we show the results of the adiabatic collapse test with this
time integration scheme. This test is one of the standard benchmark tests of
self-gravitating fluid \citep{Evrard1988}. A number of simulation codes showed
results of this test \citep[e.g.,][]{HernquistKatz1989, SteinmetzMueller1993,
Springel2005}. The typical error of the total energy in this test is less than
$1$\%.

The initial condition of this test was generated as follows.
First, we prepared gas particles located on uniform grid points. Then, by
stretching their relative distances from the center of the coordinates, we made
the gas distribution with a radial density profile of $1/r$, where $r$ is
the distance from the center of the coordinates. The gas particles in $r<1$ are
adopted to make a gas sphere. We assumed that they do not have a bulk velocity
at the beginning of the simulation.  The total mass and the gravitational
constant are unity. The specific initial thermal energy is $0.05$ and $\gamma =
5/3$.  Since the initial kinetic energy is zero and the thermal energy is less
than the absolute value of the potential energy, this gas sphere collapses due
to self-gravity and eventually reaches a state of equilibrium. Here, we
represented this gas sphere with the 61432 equal-mass SPH particles. The
gravitational softening length was set to $0.0034$.

We carried out four runs. Two are DISPH and SSPH, which were used in this paper.
In these runs, the kernel size was determined so that the neighbor number is in
the range of $128\pm8$. However, it is pointed out that the use of the constant
neighbor number often makes sudden change of the interaction radius, resulting
in the fluctuations of physical quantities. Thus, the other two runs adopted the
smoothed neighbor number method for the evaluation of kernel size
\citep{SpringelHernquist2002}.  Here, both DISPH and SSPH were used.  We express
our fiducial cases as DISPH and SSPH, and the other two cases with the smoothed
neighbor number method as DISPH(S) and SSPH(S).  

The Wendland kernel C4 was adopted. The self-gravity was solved by
using the tree with GRAPE method. The opening angle was set to $0.5$. Time step
for each particle is determined following \cite{SaitohMakino2010}.

\begin{figure}[htb]
\centering
\epsscale{1.0}
\plotone{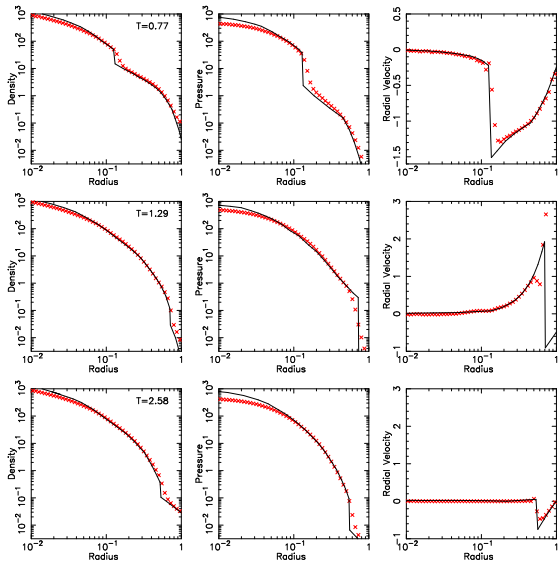}
\caption{Radial profiles of three different epochs.  From left to right, radial
profiles of density, pressure and radial velocity are shown. 
According to \cite{SteinmetzMueller1993}, both the density and the pressure are
normalized by a factor of $3/4\pi$.  Red symbols represent the averaged
quantities of the DISPH run with the bin width of $0.04$ dex. Black curves
represent the result of a 1D PPM calculation. These curves are obtained from
figure 7 in \cite{SteinmetzMueller1993}.  The DISPH run with the Wendland kernel
C4 and $N_{\rm nb} = 128 \pm 8$ is adopted.
\label{fig:3DCollapse:evolution}
}
\end{figure}

Figure \ref{fig:3DCollapse:evolution} shows the evolution of radial profiles of
density, pressure, and radial velocity of the DISPH run. Three representative
epochs are adopted. Except for the central region ($r<0.1$), our result is
comparable with a 1D PPM result.  The other three cases, SSPH, DISPH(S), and
SSPH(S), show almost identical results.

\begin{figure}[htb]
\centering
\epsscale{1.0}
\plotone{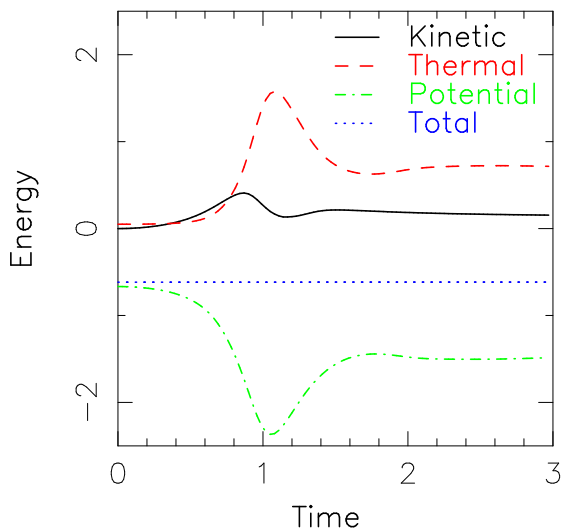}
\caption{Kinetic, thermal, potential, and total energies as a function of time.
The DISPH run with the Wendland kernel C4 and $N_{\rm nb} = 128 \pm 8$ is
adopted.
\label{fig:3DCollapse:all}
}
\end{figure}

In figure \ref{fig:3DCollapse:all}, we show time evolutions of the kinetic,
thermal, potential and total energies. Again, we show the result of the DISPH run.
The overall evolution is comparable with that obtained in previous studies and
the other three cases show identical results.

\begin{figure}[htb]
\centering
\epsscale{1.0}
\plotone{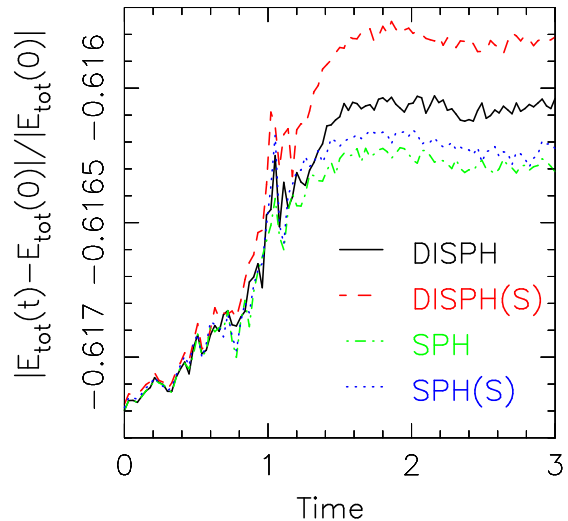}
\caption{Total energies of four models as a function of time.
\label{fig:3DCollapse:total}
}
\end{figure}

Figure \ref{fig:3DCollapse:total} represents time evolutions of the total
energies in four runs.{\footnote{We use SPH and SPH(S) in order to
denote the results of SSPH and SSPH(S) in this figure.}}
These total energies change significantly at time $\sim
1$ when the gas sphere collapses, while they keep almost a constant value at
other times.  We define the energy error as follows:
\begin{equation}
E_{\rm diff} = \frac{|E_{\rm tot}(3)-E_{\rm tot}(0)|}{|E_{\rm tot}(0)|},
\end{equation}
where $E_{\rm tot}(0)$ and $E_{\rm tot}(3)$ are the total energy of a system at
time $0$ and $3$, respectively.  Energy errors evaluated by the above equation
are 0.19\% for DISPH, 0.22\% for SSPH, 0.14\% for DISPH(S), and 0.16\% for
SSPH(S), respectively. Thus our time integration scheme works well.  The fact
that there is no significant difference among them indicates that the choices of
the definition of the neighbor numbers and SPH types are not crucial for the
energy conservation in our simulations.

\section{Definition of the Entropy Core Radius} \label{sec:method:Core}

In order to compare sizes of entropy core and their time evolutions among
different simulations, we need a methodology to evaluate a core size from an
entropy profile.  As is shown in the text, the evolutions of core size and
entropy are highly time-variable.  In order to follow this dynamically evolving
structure, we adopt the technique used in the field of stellar cluster
simulation.  Below, we describe our method to define the entropy core.

First, we calculate the radially averaged entropy profile; It is expressed as
$s(R_{\rm d})$
{\footnote{We prepare 15 bins which covers from $0.01~{\rm Mpc}$ to $10~{\rm Mpc}$ from the
density center. The constant interval with the logarithmic scale is used.  We
calculate the entropy of each particle and add it to a bin which covers the
position of the particle. We obtain the radial entropy profile of the cluster by
taking the average of entropy in each bin.  The averaged entropy at the given
$R_{\rm d}$ is calculated by the linear interpolation of the binning values.  If
$R_{\rm d} < 0.01~{\rm Mpc}$, we use $s(R_{\rm d}=0.01~{\rm Mpc})$ as the
averaged entropy at the given $R_{\rm d}$.}}, where
$R_{\rm d}$ is the distance from the density center, $\boldsymbol x_{\rm c}$.
We, then, pick up particles of which entropy is in the range of $s(R_{\rm
d})\pm0.5$ and evaluate the core radius using these particles. 
The reason why we sampled particles from the narrow region is to exclude the
contamination of small, low-entropy clumps (see also figure
\ref{fig:Entropy:Merger:Minor}).  Without this process, we were unable to obtain
a reasonable size of the entropy core.  

The following equation is used in order to evaluate the core radius:
\begin{equation}
R_{\rm {core}} = \frac{\sum_{k} \exp(-4 s_k) 
R_{{\rm d},k}}{\sum_k
\exp(-4 s_k)}, \label{eq:EntropyCore}
\end{equation}
where $s_k$ is the entropy of particle $k$ and $R_{{\rm d},k} = |\boldsymbol
x_k-\boldsymbol x_{\rm c}|$. The index $k$ runs particles of which entropy are
in the range of $s(R_{\rm d})\pm0.5$ and $R_{\rm d} < R_{200}$. We use a very
strong power of the inverse of the entropy as the weight function to determine
the entropy core.  The outer part of the entropy
profile of clusters of galaxies, typically in the range of $0.2 \leq R/R_{200}
\leq 1.0$, follows $\sim (R/R_{200})^{1.1\sim1.2}$ \citep{TozziNorman2001,
Voit+2005}. Therefore, a weight which is stronger than $1/s_j^{1.1\sim1.2}$ is
suitable in order to evaluate of the core.  After intense tests, we found that
the weight, $\exp(-4 s_j)=1/\exp(s_j)^4$, is suitable in order to evaluate the
core radius of the radial entropy profile.

\section{The Clump Finder} \label{sec:method:Subfind}

We adopt the subfind algorithm \citep{Springel+2001} in order to extract self-bound
clumps in the Santa Barbara cluster. The procedure is as follows.  First, we
calculate the density of all particles within $R_{\rm 200}$ using Eq
\eqref{eq:SSPH}. The density of gas particles is evaluated by counting the
contribution of gas particles while that of DM particles is calculated
by counting the contribution of DM particles.  The density of each particle is
evaluated by the contribution from the nearest $128\pm8$ particles.  Here, we
refer these nearest particles ``local particles''. After the density evaluation,
we sort all of particles in the descending order of the density.

We, then, identify isolate clumps following the density distribution.  We
repeatedly compare the densities of particles with those in the densities of
their local particles from the highest to the lowest density particles and 
apply the following operation. Here, we focus a particle $i$.  If the particle
$i$ is the highest density particle in its local particles, we set the particle
$i$ as the peak of an isolate clump (the core of a new clump). If the particle
$i$ is the second highest density particle or it has two higher density
particles which are in the same clump, we add the particle $i$ to the clump
which includes the highest density particle.  If the particle $i$ has more than
two higher density particles and the two highest density particles are in
different clumps, we tag these two clumps as {\it the subfind clump candidates}
since the particle $i$ is at the saddle point which separate two clumps (two
density peaks).  Then, we merge these clumps with a single larger clump and
attach particle $i$ to this clump. The last operation is necessary in order to
extract subfind clump candidates which are located at the envelop of this merged
clump.

As the final step, we remove unbound particles from the subfind clump candidates
iteratively.  We first calculate the total energy (the self-gravitational
potential and the kinematic energy where the mean velocity of each clump is set
to zero) of each particle in a clump. Then if the clump includes particles of
which total energy is larger than zero, we remove these particles from the
clump.  We do not remove more that $10\%$ of the particles at once so that we
can carry out this process stably.  We continue this procedure until the clump
consists only of the bound particles.  We adopt clumps which contains more than
64 bound particles.  We apply this procedure to all clumps and get a full
catalog of subfind clumps.

\section{Mass Functions of Substructures} \label{sec:results:MassFunction}

In this section we investigate the mass functions of substructures. The
subclumps are extracted using the subfind algorithm (see \S
\ref{sec:method:Subfind}).

\begin{figure}
\centering
\epsscale{1.0}
\plotone{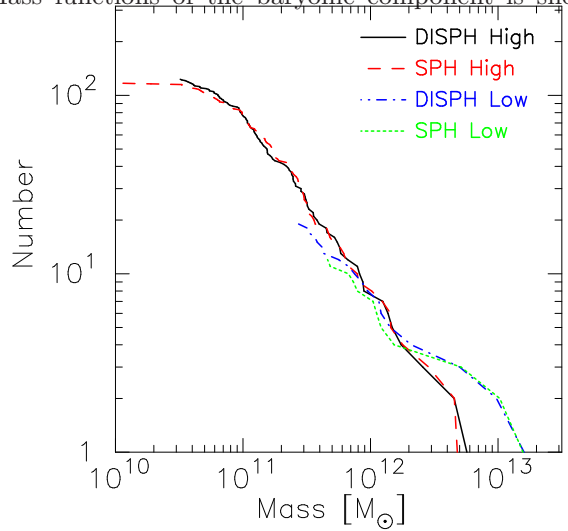}
\caption{
Cumulative mass functions of substructures found in the cluster with different
resolutions and different schemes.
\label{fig:MassFunction}
}
\end{figure}

Figure \ref{fig:MassFunction} shows mass functions of substructures.  Mass
functions of the runs with the same resolutions have almost identical profiles.
All of mass functions asymptotically follow the power law profile with the index
of $\sim -1$.  It is not surprising that there is no impact on substructure mass
functions with different schemes, since the amount of the baryonic component is
$\sim 10\%$ of the total mass.

\begin{figure}
\centering
\epsscale{1.0}
\plotone{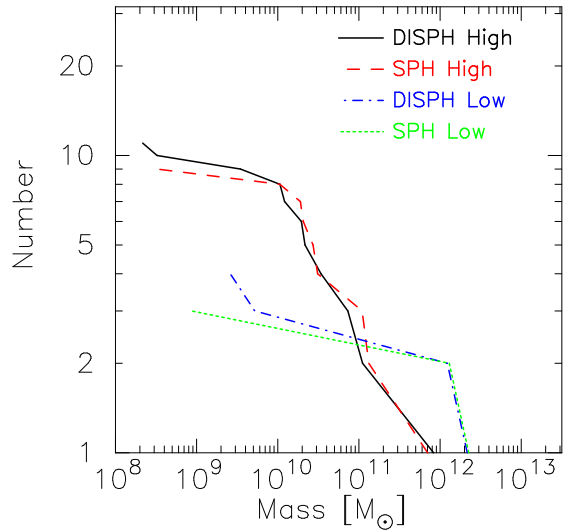}
\caption{The same as figure \ref{fig:MassFunction}, but for baryons found in
substructures.
\label{fig:GasMassFunction}
}
\end{figure}

Mass functions of the baryonic component is shown in figure
\ref{fig:GasMassFunction}.  Not all of substructures hold the baryonic component
due to the stripping. Thus, the number of baryonic substructures is very small
compared to that of the total substructures.  Even though the evolution of the
gas clumps in the central area of the cluster is quite different for two runs,
this difference is little or no effect on the mass function. This is probably
because in either case that lifetime of gas clump in the central area is short.

Our mass functions do not show convergence with different mass resolutions.
We have checked mass functions with mixed resolutions and found that, at least
toward the high mass end, the functional form of mass functions depends on the mass
resolution. Thus, the discrepancy between high-mass resolution runs and low-mass
resolution runs comes from the non-linear effect induced by the smallest scale
density fluctuations.

\end{appendix}

\end{document}